\documentclass{biometrika}

\usepackage{amsmath}
\usepackage{url}
\usepackage{newtxtext}
\usepackage[subscriptcorrection]{newtxmath}
\usepackage{graphicx}
\graphicspath{{./art/}}
\newcommand{\figurealt}[1]{%
  \par\noindent\makebox[\linewidth][c]{%
    \parbox[t]{\figwidth}{\figcaptionfont\centering\noindent Alt text: #1\par}}%
  \par}
\newcommand{\tablealt}[1]{%
  \par\begingroup
  \centering
  \advance\leftskip by \tabledim
  \advance\rightskip by \tabledim
  \noindent Alt text: #1\par
  \endgroup}

\usepackage[plain,noend]{algorithm2e}

\makeatletter
\renewcommand{\algocf@captiontext}[2]{%
  #1\algocf@typo. \AlCapFnt{}#2}

\def\@algocf@capt@plain{top}
\renewcommand{\algocf@makecaption}[2]{%
  \addtolength{\hsize}{\algomargin}%
  \sbox\@tempboxa{\algocf@captiontext{#1}{#2}}%
  \ifdim\wd\@tempboxa>\hsize
    \hskip .5\algomargin
    \parbox[t]{\hsize}{\algocf@captiontext{#1}{#2}}%
  \else
    \global\@minipagefalse
    \hbox to\hsize{\box\@tempboxa}%
  \fi
  \addtolength{\hsize}{-\algomargin}%
}
\makeatother

\makeatletter
\let\biometrikaOriginalPlain\ps@plain
\def\ps@plain{%
  \biometrikaOriginalPlain
  \let\@oddhead\@empty
  \let\@evenhead\@empty
}
\makeatother
\begin{document}

\jname{}
\markboth{Zihan Xu}{Exact local queries for pseudo-marginal Markov chains}

\title{Exact local queries for pseudo-marginal Markov chains}
\author{Zihan Xu}
\affil{School of Mathematics and Statistics, Qingdao University, Qingdao, 266071, China
\email{xuzihan1@qdu.edu.cn}}
\maketitle

\begin{abstract}
Local queries of a Markov chain can depend on past updates outside the queried components.
For pseudo-marginal chains with Gaussian graph priors, we construct decision certificates from finite historical parameter ranges to settle updates before all neighbouring records are resolved.
Under shared random inputs and exact arithmetic, the executor recovers the reference parameters, auxiliary variables and likelihood estimates for any prespecified finite query sequence.
We bound conditional expected work, including likelihood evaluation, historical range construction and execution management.
On specified path and grid families, we fix per-node history lengths and all other parameters in the work bounds.
As graph size grows, expected incremental work stays bounded for certificate execution and grows linearly for the specified cached pointwise evaluator answering the same query.
In controlled experiments, certificate execution is faster than pointwise and parallel Picard execution for sparse queries, whereas Picard is faster than certificate execution for full queries.
We apply the method to positron emission tomography and Gaussian process classification of Landmine data.
For a fixed numerical tomography posterior, certificate execution completes all queries of regional mean distribution volume within budget, while full execution completes none from the same initial cache.
In Landmine, posterior summaries of the difference in log length scales between two sites agree across pseudo-marginal and ancillary augmentation samplers within estimated Monte Carlo uncertainty.
Certificate execution resolves almost every scheduled update over the full query sequence.
\end{abstract}


\section{Introduction}
Bayesian inference often concerns functions of a small subset of the
parameters in a joint model.
Examples include regional averages in positron emission tomography and
contrasts between covariance length scales at different sites.
In Markov chain Monte Carlo (MCMC), these outputs can depend on earlier
updates to parameters outside the queried subset.
Such updates can be expensive when likelihood evaluation requires
integration over latent variables, as in Gaussian process classification
\citep{FilipponeGirolami2014}.
Pseudo-marginal methods use nonnegative, unbiased likelihood estimates
to construct a chain whose invariant distribution has the required
posterior marginal \citep{AndrieuRoberts2009}.
We study how to recover specified joint outputs from a prescribed reference
chain while reducing the historical computation needed to determine them.

One line of work concerns computation within individual updates.
In pseudo-marginal methods, likelihood estimation involves a trade-off
between computational cost and chain mixing
\citep{DoucetEtAl2015,SherlockEtAl2015}.
Correlating auxiliary variables can reduce noise in estimates of likelihood
ratios \citep{DeligiannidisEtAl2018}.
Delayed acceptance uses a cheap approximation to screen proposals before
an expensive evaluation \citep{SherlockThieryGolightly2021}.
Subsampling methods reduce the number of likelihood terms evaluated per
iteration \citep{BardenetDoucetHolmes2017}.
Some use control variates and bounds to preserve the exact posterior as
the invariant distribution \citep{CornishEtAl2019,PradoNemethSherlock2026}.

Other methods reorganize the simulation of a reference chain.
Predictive prefetching evaluates possible future steps in parallel while preserving the serial trajectory \citep{AngelinoEtAl2014}.
Picard methods formulate simulation with fixed random inputs as a fixed-point problem over trajectories \citep{GrazziZanella2026}.
For discrete spin systems, \citet{FengEtAl2025CTTP} introduced coupling toward the past to recover local states by tracing historical dependencies and caching resolved states.
\citet{LiuWangYin2026} develop local sampling procedures that also use partial information about unresolved states.

We develop an exact local query method for pseudo-marginal chains with local likelihoods and a Gaussian graph prior.
A complete record contains a parameter, its auxiliary variables and its likelihood estimate.
For additive random walks and independent proposals, resolved states and fixed proposal inputs over a finite history give finite, computable ranges for unresolved historical parameters.
A resolved parameter anchors ranges for later updates at the same node.
With the old and candidate records fixed and the candidate likelihood estimate positive, the log acceptance ratio is affine in the neighbouring parameters.
The ranges bound this ratio and can determine the update from its original uniform threshold before all neighbouring records are resolved.
This comparison uses the principle of retrospective sampling \citep{PapaspiliopoulosRoberts2008,ChanialidisEtAl2014,LiSraJegelka2016}.

A recursive executor uses these bounds as decision certificates for joint queries and caches complete records.
We prove that the executor terminates and returns the same complete records as full reference execution for any prespecified finite query collection and request order (Theorem~\ref{thm:query-correctness}).
The result assumes shared initial records, update schedule and random inputs, exact arithmetic, and finite time to generate and evaluate each candidate.
Each scheduled candidate likelihood estimate is computed at most once.
The two executions also return the same statistical outputs when the additional random inputs used to generate those outputs are shared.

For independent standard Gaussian proposals on bounded-degree graphs, a contraction condition bounds expected incremental work by a constant times the number of queries, uniformly in graph size (Corollary~\ref{cor:bounded-degree-work}).
With these proposals, the path construction in Theorem~\ref{thm:path-block-work} gives a geometric tail bound on the number of resolved blocks at every fixed finite coupling strength.
Both results allow candidate costs to depend on the proposal inputs governing propagation, under a common integrable cost bound.
For the specified path and grid schedules, expected incremental work is bounded for certificate execution and linear in graph size for the specified cached pointwise evaluator (Corollary~\ref{cor:pointwise-separation}).
The comparison uses the same single-record query, initial-record cache and random inputs, with almost surely positive candidate likelihood estimates.
These size comparisons fix the model parameters, input laws, per-node history lengths, initial parameter bounds, operation costs and common cost bound.
We also bound remaining work after a completed request by conditioning on proposal inputs and revealed thresholds (Theorem~\ref{thm:conditional-work}).
The accounting includes record evaluation, input and range generation, and execution management; preparation and statistical output are charged separately from incremental work.

Controlled experiments measure the work of block resolution and compare
execution strategies for the same queries.
For the two-node queries studied, certificate execution requires fewer
candidate likelihood evaluations than cached pointwise execution and takes
less time than either pointwise or parallel Picard execution.
In the full-graph comparisons, parallel Picard execution is faster.

In positron emission tomography, we infer the posterior distribution of mean distribution volume in the left insula.
We extend the certificate construction to binary model labels coupled across voxels by a Potts prior.
Fixed quadrature and binning define the numerical posterior.
From the same initial cache, certificate execution completes all prescribed regional queries within the time budget.
Full execution reaches the budget before completing its first query.

For Landmine classification data, we infer the difference between two sites' log covariance length scales.
The pseudo-marginal sampler uses unbiased likelihood estimates and jointly updates each site's log length scale and log amplitude.
Early decisions are frequent, and almost every scheduled update is resolved over the full query sequence.
Later requests can require records bypassed by earlier certificates.
Posterior summaries agree within estimated Monte Carlo uncertainty with those from an ancillary augmentation sampler targeting the same posterior.

\section{Reference chains and queries}
\label{sec:reference}

\subsection{Target distribution and reference chain}
\label{sec:reference-chain}

We develop the method for a graph model with scalar node parameters.
Let $G=(V,E)$ be a finite undirected graph with node set
$V=\{1,\ldots,n\}$ and edge set $E$.
At node $j\in V$, the observed data $y_j$ have likelihood
$\ell_j(\theta_j)$ for a scalar parameter $\theta_j\in\mathbb{R}$.
For $\theta=(\theta_1,\ldots,\theta_n)\in\mathbb{R}^n$, consider the
posterior density
\begin{equation}
\label{eq:graph-posterior}
\pi(\theta)
=\frac{1}{Z}\left\{\prod_{j=1}^{n}\ell_j(\theta_j)\right\}
\exp\left\{
-\frac12\sum_{j=1}^{n}a_j\theta_j^2
-\frac12\sum_{\{j,k\}\in E}w_{jk}(\theta_j-\theta_k)^2
\right\},
\end{equation}
where $a_j>0$ are fixed node coefficients, $w_{jk}>0$ are fixed edge
weights, and $Z$ is the positive, finite normalizing constant.
The likelihood terms are local, and the prior couples neighbouring parameters.

At node $j$, let $m_j(\mathrm{d}u_j\mid\theta_j)$ denote the simulation
law of an auxiliary variable $u_j\in\mathcal{U}_j$.
Assume that the finite nonnegative likelihood estimator
$\widehat{\ell}_j(\theta_j,u_j)$ satisfies
\[
\int_{\mathcal{U}_j}
\widehat{\ell}_j(\theta_j,u_j)\,
m_j(\mathrm{d}u_j\mid\theta_j)
=\ell_j(\theta_j),
\qquad \theta_j\in\mathbb{R}.
\]
Write $u=(u_1,\ldots,u_n)$, and let $g(\theta)$ denote the exponential
factor in \eqref{eq:graph-posterior}.
The corresponding pseudo-marginal target \citep{AndrieuRoberts2009} is
\[
\widetilde{\pi}(\mathrm{d}\theta,\mathrm{d}u)
=\frac{g(\theta)}{Z}
\left\{
\prod_{j=1}^{n}
\widehat{\ell}_j(\theta_j,u_j)\,
m_j(\mathrm{d}u_j\mid\theta_j)
\right\}\mathrm{d}\theta.
\]
Integrating out $u$ recovers the posterior in
\eqref{eq:graph-posterior}.
The complete record at node $j$ is
$R_j=(\theta_j,u_j,\widehat{\ell}_j(\theta_j,u_j))$.
We retain $R_j$ whenever a proposal is rejected.
Deterministic likelihood evaluation is included by taking
$\widehat{\ell}_j(\theta_j,u_j)=\ell_j(\theta_j)$.

Fix initial records $R_j^{(0)}$, $j\in V$, with positive likelihood
estimates and a schedule of $T$ node updates $j_1,\ldots,j_T\in V$.
Write $R_j^{(t)}$ for the record at node $j$ after $t$ updates.
The parameter proposal at node $j$ has a fixed, positive, finite
density $q_j(x,y)$ for a move from $x$ to $y$.
At update $t$, set $j=j_t$ and let $(\theta,u)$ be the state after
$t-1$ updates.
Draw $\vartheta\sim q_j(\theta_j,\cdot)$ and then
$v\sim m_j(\cdot\mid\vartheta)$.
The acceptance probability is
\begin{equation}
\label{eq:reference-acceptance}
\alpha_t
=\min\left\{
1,
\frac{
g(\theta^{j\leftarrow\vartheta})\,
\widehat{\ell}_j(\vartheta,v)\,q_j(\vartheta,\theta_j)
}{
g(\theta)\,\widehat{\ell}_j(\theta_j,u_j)\,
q_j(\theta_j,\vartheta)
}
\right\},
\end{equation}
where $\theta^{j\leftarrow\vartheta}$ is obtained from $\theta$ by
replacing its $j$th coordinate with $\vartheta$.
Let $U_1,\ldots,U_T$ be independent uniform thresholds on $(0,1]$,
independent of the random inputs used to generate proposals.
Set $R_j^{(t)}=(\vartheta,v,\widehat{\ell}_j(\vartheta,v))$ if
$U_t<\alpha_t$, and $R_j^{(t)}=R_j^{(t-1)}$ otherwise.
All other node records remain unchanged.
A zero likelihood estimate gives $\alpha_t=0$.
Each node kernel preserves $\widetilde{\pi}$.
Fixing the underlying random inputs for proposal generation and the
thresholds $U_t$ determines the reference trajectory.
This fixed-input representation also underlies the Picard simulation
of Markov chains \citep{GrazziZanella2026}.

\subsection{Queries and computational work}
\label{sec:queries-work}

Fix a finite set $Q\subseteq V\times\{0,\ldots,T\}$ of requested
node and time pairs and an order for issuing these requests,
independently of the random inputs of the reference chain.
The requested output is the indexed collection
\begin{equation}
\label{eq:joint-query-output}
\mathcal{R}_Q=\bigl(R_j^{(t)}\bigr)_{(j,t)\in Q}.
\end{equation}
All records in this collection come from the same realization of
the reference chain.
Let $S\subseteq V$ be the set of nodes occurring in $Q$, and write
$r=|S|$.
If these nodes are queried at the same $K$ prescribed times
$0\le t_1<\cdots<t_K\le T$, then
$Q=S\times\{t_1,\ldots,t_K\}$ and $|Q|=rK$.

For the common query times above, fix a measurable real-valued
output function $h$ and write
$Y_k=h\bigl((R_j^{(t_k)})_{j\in S},\xi_k\bigr)$,
$k=1,\ldots,K$.
Conditional output draws use auxiliary inputs $\xi_k$ with
prescribed laws, independent across $k$ and of the reference chain.
We share these inputs across execution methods and omit them for
deterministic outputs.
Choices of $h$ include a parameter contrast, a regional conditional
mean for mean estimation and a regional average of conditional
draws for quantile estimation.
Let $\mathcal{I}\subseteq\{1,\ldots,K\}$ be a nonempty,
prespecified set of retained indices.
The empirical mean and positive fraction are
\[
\overline{Y}
=\frac{1}{|\mathcal{I}|}\sum_{k\in\mathcal{I}}Y_k,
\qquad
\widehat{p}_{+}
=\frac{1}{|\mathcal{I}|}\sum_{k\in\mathcal{I}}
\mathbf{1}\{Y_k>0\},
\]
where $\mathbf{1}\{\cdot\}$ denotes the indicator function.
Empirical quantiles use the retained outputs and a fixed
interpolation rule.

Fix an additive cost model.
Let $W_{\mathrm{setup}}$ denote preparation work and
$W_{\mathrm{out}}(Q)$ the work of assembling the requested records
and computing their statistical outputs, including any conditional draws.
Starting from the prepared state, let $W_{\mathrm{back}}(Q)$ be
the cost of new likelihood computations and $W_{\mathrm{manage}}(Q)$
the remaining work needed to resolve the records.
We define the incremental and total work by
\begin{equation}
\label{eq:query-work}
\begin{split}
W_{\mathrm{inc}}(Q)
&=W_{\mathrm{back}}(Q)+W_{\mathrm{manage}}(Q),\\
W_{\mathrm{total}}(Q)
&=W_{\mathrm{setup}}+W_{\mathrm{inc}}(Q)+W_{\mathrm{out}}(Q).
\end{split}
\end{equation}
Preparation includes data and graph processing, initial records
and caches, and any precomputed random inputs or indices.
Management includes candidate generation, range calculations,
cache access and scheduling; each operation is charged once.
Comparisons from an existing cache identify that cache and
account for its preparation separately.
We report likelihood evaluation counts and measured elapsed times
separately from these work measures.

\section{Execution with decision certificates}
\label{sec:certificate-execution}

\subsection{Historical ranges and decision certificates}
\label{sec:history-certificates}

A record is \emph{resolved} when all its components in the fixed
reference trajectory have been determined.
At update $t$, define
\[
p_k(t)=\max\left(
\{0\}\cup\{s\in\{1,\ldots,t-1\}:j_s=k\}
\right),
\qquad k\in V,
\]
where $p_k(t)$ is the last scheduled update time of node $k$ strictly
before $t$, and zero denotes its initial record.
For $j=j_t$, the old record is $R_j^{(p_j(t))}$.
We resolve this record first and use it with the fixed proposal
inputs to generate and evaluate the candidate.
The acceptance ratio in \eqref{eq:reference-acceptance} uses the
parameter components of $R_k^{(p_k(t))}$ for neighbouring nodes $k$.
We enclose these parameters in intervals to test whether the update
can be decided before all neighbouring records are resolved.

Bounding chains propagate sets of possible states under shared random updates \citep{Huber2004BoundingChains}.
\citet{MurdochGreen1998} use bounds on Metropolis--Hastings acceptance probabilities to couple chains on continuous state spaces.
For additive random walks and independent proposals, we construct explicit historical ranges from fixed proposal inputs and a resolved record.
Start from the latest resolved record of node $k$ at or before
$p_k(t)$.
Write $x_0$ for its parameter and $s_1<\cdots<s_m\le p_k(t)$ for
the subsequent updates of node $k$.
For the random walk, $\delta_i$ is the fixed proposal increment at
$s_i$; for the independent proposal, $\eta_i$ is the fixed candidate value.
For $m\ge1$, the parameter component of $R_k^{(p_k(t))}$ lies in
\begin{equation}
\label{eq:historical-ranges}
H_m=
\begin{cases}
\displaystyle
\left[x_0+\sum_{i=1}^{m}\min(0,\delta_i),\;
      x_0+\sum_{i=1}^{m}\max(0,\delta_i)\right],
& \text{random walk},\\
\bigl[\min\{x_0,\eta_1,\ldots,\eta_m\},\;
      \max\{x_0,\eta_1,\ldots,\eta_m\}\bigr],
& \text{independent proposal}.
\end{cases}
\end{equation}
For $m=0$, take $H_0=\{x_0\}$.
Each update either keeps its previous parameter or takes its candidate,
which gives these bounds by induction.
Recomputing the ranges for the same later times from a newly resolved
record at node $k$ can only tighten them.

Write the resolved old record as
$(x,u_j,\widehat{\ell}_j(x,u_j))$ and the candidate as
$(\vartheta,v,\widehat{\ell}_j(\vartheta,v))$.
Assume that $\widehat{\ell}_j(\vartheta,v)>0$.
Let $\mathcal{N}_j=\{k:\{j,k\}\in E\}$, and write $z_k$ for the
parameter component of $R_k^{(p_k(t))}$.
Expanding the graph prior gives the log Metropolis--Hastings ratio
\[
\begin{aligned}
\Lambda_t
&=C_t+\sum_{k\in\mathcal{N}_j}c_{tk}z_k,
\qquad c_{tk}=w_{jk}(\vartheta-x),\\
C_t
&=\log\frac{\widehat{\ell}_j(\vartheta,v)q_j(\vartheta,x)}
{\widehat{\ell}_j(x,u_j)q_j(x,\vartheta)}
-\frac12\left(a_j+\sum_{k\in\mathcal{N}_j}w_{jk}\right)
(\vartheta^2-x^2).
\end{aligned}
\]
All terms in $C_t$ are known at this point.
For each neighbour, obtain an interval
$z_k\in[\underline{z}_k,\overline{z}_k]$
from \eqref{eq:historical-ranges}, using a singleton for a resolved record.
The lower and upper bounds are
\begin{equation}
\label{eq:log-ratio-bounds}
\begin{aligned}
\underline{\Lambda}_t
&=C_t+\sum_{k\in\mathcal{N}_j}
\min\{c_{tk}\underline{z}_k,c_{tk}\overline{z}_k\},\\
\overline{\Lambda}_t
&=C_t+\sum_{k\in\mathcal{N}_j}
\max\{c_{tk}\underline{z}_k,c_{tk}\overline{z}_k\}.
\end{aligned}
\end{equation}
Thus $\underline{\Lambda}_t\le\Lambda_t\le\overline{\Lambda}_t$, with
\[
\overline{\Lambda}_t-\underline{\Lambda}_t
=|\vartheta-x|\sum_{k\in\mathcal{N}_j}
w_{jk}(\overline{z}_k-\underline{z}_k).
\]
Resolved neighbours contribute zero to this width.

\begin{figure}[tb]
\centering
\setlength{\figwidth}{\linewidth}
\includegraphics[width=\figwidth]{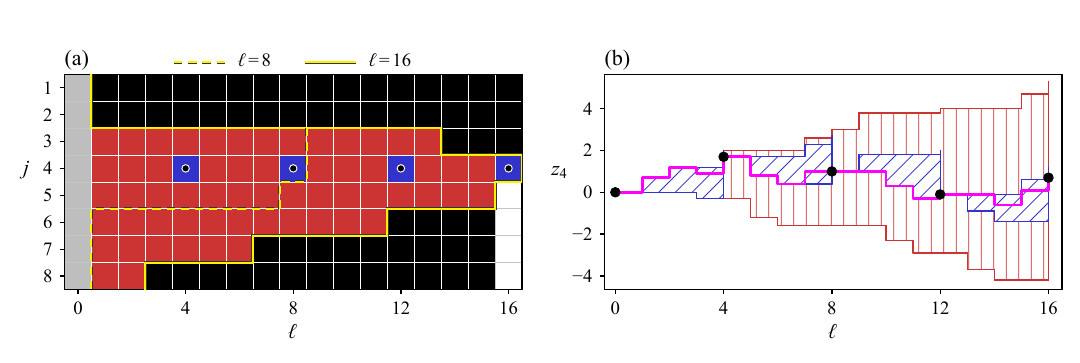}
\caption{Historical resolution and legal ranges on an eight-node path with
fixed additive random-walk inputs. Each round $\ell$ updates nodes
$1,\ldots,8$ in order; node 4 is queried after its updates in rounds
4, 8, 12 and 16. Initial parameters are zero, node precisions are
$1/2$, edge weights are $1/4$, and likelihoods are one.
(a) Records after the round-16 query: blue dotted cells are queried,
red cells otherwise resolved, black cells unresolved, grey cells initial,
and white cells after the current query. Dashed and solid yellow
frontiers correspond to rounds 8 and 16.
(b) Ranges for $z_4$, the parameter after node 4's round-$\ell$ update.
This panel follows the four queries in order.
Red vertical hatching shows ranges from the initial anchor; blue diagonal hatching shows ranges from the latest resolved anchor.
At each query, the blue range collapses to the returned parameter value.
The magenta line shows the full reference path for comparison, with black dots at the initial and queried states.}
\label{fig:decision-certificate}
\figurealt{Panel (a) maps queried, resolved and unresolved records across nodes and update rounds. Panel (b) shows how resolved historical anchors narrow the parameter ranges. At each query, the ranges collapse to the reference path.}
\end{figure}

The acceptance test refines bounds against a fixed uniform threshold,
as in retrospective sampling \citep{ChanialidisEtAl2014,LiSraJegelka2016}.
A zero candidate estimate is rejected immediately.
For a positive estimate, compare the bounds in \eqref{eq:log-ratio-bounds}
with the original threshold $U_t$:
\begin{equation}
\label{eq:decision-certificates}
\begin{cases}
\text{accept},
& \log U_t<\min\{0,\underline{\Lambda}_t\},\\
\text{reject},
& \log U_t\ge\min\{0,\overline{\Lambda}_t\},\\
\text{resolve a neighbour},
& \text{otherwise}.
\end{cases}
\end{equation}
Each of the first two conditions is a \emph{decision certificate}:
it fixes the reference decision for all neighbour parameter values
in the current intervals.
If neither holds, resolve an unresolved record $R_k^{(p_k(t))}$,
$k\in\mathcal{N}_j$, and repeat the comparison with updated bounds,
using the same old record, candidate and threshold.
Every rejection retains the complete old record.
With one neighbour, for example, take $C_t=-1/2$, $c_{tk}=1/4$
and $z_k\in[-1,1]$.
The bounds are $[-3/4,-1/4]$, so $\log U_t=-1$ certifies
acceptance before that neighbour is resolved. Figure~\ref{fig:decision-certificate} illustrates two effects of resolving
queries: the cached history grows, and new anchors tighten subsequent
parameter ranges.

\subsection{Query execution and correctness}
\label{sec:query-correctness}

Algorithm~\ref{alg:query-execution} applies the decision certificates to the requests in $Q$ in their prescribed order.
It uses recursive evaluation and a shared cache of complete records, following the structure of local Gibbs sampling \citep{FengEtAl2025CTTP,LiuWangYin2026}.
An explicit stack implements the recursion.

\begin{algorithm}[tbp]
\caption{Query execution with decision certificates.}
\label{alg:query-execution}
\SetKwProg{Fn}{Procedure}{:}{}
\SetKwFunction{Resolve}{Resolve}
\LinesNumbered
\KwIn{The reference specification in \S\ref{sec:reference-chain},
fixed random inputs, and requests $Q$ with their prescribed order.}
\KwOut{The indexed collection $\mathcal{R}_Q$.}
Initialize the shared record cache $\mathcal{C}$ with
$\mathcal{C}[j,0]=R_j^{(0)}$, $j\in V$, and initialize
$\mathcal{R}_Q$ as empty\;
\Fn{\Resolve{$j,\tau$}}{
  $t\gets\max\bigl(\{0\}\cup
    \{s\in\{1,\ldots,\tau\}:j_s=j\}\bigr)$\;
  \lIf{$(j,t)$ is cached}{\KwRet $\mathcal{C}[j,t]$}
  $R^-\gets{}$ \Resolve{$j,p_j(t)$}\;
  Generate and retain the candidate record
  $R^+=(\vartheta,v,\widehat{\ell}_j(\vartheta,v))$
  from $R^-$ and the fixed inputs at $t$, evaluating it once\;
  $R\gets R^-$\;
  \If{$\widehat{\ell}_j(\vartheta,v)>0$}{
    \While{true}{
      Form intervals for the parameter components of
      $R_k^{(p_k(t))}$, $k\in\mathcal{N}_j$, by \eqref{eq:historical-ranges}, using singletons for
      cached records; compute
      $[\underline{\Lambda}_t,\overline{\Lambda}_t]$
      by \eqref{eq:log-ratio-bounds}\;
      \lIf{$\log U_t<\min\{0,\underline{\Lambda}_t\}$}{
        $R\gets R^+$; \textbf{break}}
      \lIf{$\log U_t\ge\min\{0,\overline{\Lambda}_t\}$}{
        \textbf{break}}
      Choose an unresolved neighbour $k$ whose parameter interval
      has largest width, breaking ties by the smallest node index\;
      \Resolve{$k,p_k(t)$}; retain $R^-$, $R^+$ and $U_t$
      for the next comparison\;
    }
  }
  $\mathcal{C}[j,t]\gets R$; use its parameter component as the
  new range anchor at node $j$ and discard ranges formed from its
  previous anchor\;
  \KwRet $R$\;
}
\ForEach{$(j,\tau)\in Q$ in the prescribed order}{
  Append \Resolve{$j,\tau$} to $\mathcal{R}_Q$ with index
  $(j,\tau)$\;
}
\end{algorithm}

We establish correctness for the finite reference construction in
\S\ref{sec:reference-chain}, using either of the two proposal
constructions in \eqref{eq:historical-ranges}.
All computations are in exact arithmetic, and each candidate record
is assumed to be generated and evaluated in finite time.
The reference and query executions use the same initial records,
update schedule and random inputs.

\begin{theorem}
\label{thm:query-correctness}
Under the preceding conditions, fix any query set $Q$ and prescribed
request order as in \S\ref{sec:queries-work}.
For every realization of the reference random inputs,
Algorithm~\ref{alg:query-execution} terminates and returns the indexed
collection $\mathcal{R}_Q$ in \eqref{eq:joint-query-output}.
Every cached record agrees with the corresponding complete reference
record, including its parameter, auxiliary variable and likelihood estimate.
At each scheduled update, the candidate likelihood estimate is computed
at most once.
\end{theorem}

\begin{proof}[of Theorem~\ref{thm:query-correctness}]
Initial records are cached.
Every recursive call from update $t$ requests an earlier record.
For a positive candidate estimate, each undecided comparison resolves
a previously unresolved neighbour record.
Once all neighbours are resolved, the interval in
\eqref{eq:log-ratio-bounds} is a singleton, and
\eqref{eq:decision-certificates} decides the update, including equality
at the threshold.
Zero estimates are rejected immediately in both executions.
Induction on $t$, followed by finiteness of $Q$, proves termination.
For correctness, induct over records added to the cache, starting
from the given initial records.
Resolving own predecessors first makes the cached updates at each node
a prefix of its schedule.
Thus the range for an unresolved historical record starts from an
earlier resolved anchor.
Given correct cached records, the old record and fixed inputs generate
the same complete candidate as the reference execution.
For a positive estimate, the ranges in \eqref{eq:historical-ranges}
contain the required reference parameters, so
\eqref{eq:log-ratio-bounds} encloses the reference log ratio.
Each certificate therefore gives the reference decision.
Acceptance stores the complete candidate; rejection stores the
complete old record.
This preserves cache correctness.
Since a node's record is constant between its updates, the request-time
mapping returns $R_j^{(\tau)}$ for every $(j,\tau)\in Q$.
The candidate is evaluated before the comparison loop and retained
across recursive calls; subsequent requests use the cached result.
Hence each scheduled candidate is evaluated at most once.
The outputs $Y_k$ in \S\ref{sec:queries-work} also agree when the
same $\xi_k$ are used, as do their prescribed summaries.
\end{proof}

The positron emission tomography application uses binary labels with a Potts prior.
An unresolved historical label has feasible set $\{0,1\}$, and a resolved label gives a singleton.
Supplementary Material, \S\ref{supp:pet}, gives the resulting acceptance bounds and proves agreement with the corresponding full reference execution.

The Landmine application updates two parameters at each site, with only the log length scale entering the interactions between sites.
Supplementary Material, \S\ref{supp:landmine}, derives the block acceptance bounds and establishes agreement with full reference execution.

\section{Work bounds for query execution}
\label{sec:work-bounds}

\subsection{Conditional work bounds}
\label{sec:conditional-work}

Consider Algorithm~\ref{alg:query-execution} immediately after the first
$s$ requests in $Q$ have been completed, where $0\le s\le |Q|$.
Write $\mathcal{C}_s$ for the cache at this boundary, when no recursive
call is pending.
Let $F$ collect the raw inputs used to generate parameter and auxiliary
proposals at the $T$ updates.
These inputs are independent of the acceptance thresholds
$U_1,\ldots,U_T$.
We reveal the full value of a threshold when it is first used in an
execution decision.
The next index to reveal and the decision to stop revealing thresholds
depend only on $F$ and the values already revealed.
Let $\mathcal{T}_s$ record the revealed indices and values in order,
and set $\mathcal{G}_s=\sigma(F,\mathcal{T}_s)$.
This information determines $\mathcal{C}_s$.
At this boundary, every update with a revealed threshold is cached.
An update with a zero candidate estimate is cached without revealing
its threshold.
The thresholds of uncached updates are therefore still unrevealed;
conditionally on $\mathcal{G}_s$, they are independent and uniform
on $(0,1]$.
Supplementary Material, \S\ref{supp:conditional-proof}, proves this
claim for the adaptive revelation order.
By Theorem~\ref{thm:query-correctness}, all cached records agree with
the reference chain for every compatible completion of the
unrevealed thresholds.

For each uncached update $t$, form canonical ranges for the neighbour
parameters at times $p_k(t)$.
Use a singleton when the required record is in $\mathcal{C}_s$;
otherwise apply \eqref{eq:historical-ranges} from the latest cached
anchor at or before $p_k(t)$.
With the reference old and candidate records, these ranges give
canonical log-ratio bounds
$[\underline{\Lambda}_t^{\circ},\overline{\Lambda}_t^{\circ}]$
through \eqref{eq:log-ratio-bounds} when the candidate estimate is positive.
Choose a finite, nonnegative, $\mathcal{G}_s$-measurable bound
$\omega_t$ on their width, valid for every compatible completion.
The corresponding undecided band is
$[\underline{\alpha}_t,\overline{\alpha}_t)$, where
$\underline{\alpha}_t=\min\{1,\exp(\underline{\Lambda}_t^{\circ})\}$ and
$\overline{\alpha}_t=\min\{1,\exp(\overline{\Lambda}_t^{\circ})\}$;
set $\underline{\alpha}_t=\overline{\alpha}_t=0$ for a zero estimate.
The elementary exponential bound in
\citet[Supplementary Material, Lemma~1]{GrazziZanella2026} gives
\begin{equation}
\label{eq:canonical-band-width}
\overline{\alpha}_t-\underline{\alpha}_t
\le p_t:=1-\exp(-\omega_t).
\end{equation}
List the uncached updates as $\tau_1<\cdots<\tau_m$ in the reference
schedule, where $m$ is their number.
For $r=1,\ldots,m$, the canonical endpoints at $\tau_r$ depend only
on $\mathcal{G}_s$ and $U_{\tau_1},\ldots,U_{\tau_{r-1}}$.
Set $V_{\tau_r}=(U_{\tau_r}-\underline{\alpha}_{\tau_r})\bmod 1$.
Each shift is determined before $U_{\tau_r}$ is used in this order,
so successive conditioning shows that the $V_{\tau_r}$ are
independent uniforms on $[0,1)$ given $\mathcal{G}_s$.
Thus the events
\begin{equation}
\label{eq:dominating-events}
B_t=\{V_t<p_t\},
\qquad \operatorname{pr}(B_t\mid\mathcal{G}_s)=p_t,
\end{equation}
are conditionally independent over uncached updates $t$.
A recursive call from update $t$ to an unresolved neighbour requires
$U_t\in[\underline{\alpha}_t,\overline{\alpha}_t)$, because the
execution ranges refine the canonical ranges.
By \eqref{eq:canonical-band-width}, this implies $B_t$.
The execution retains $U_t$ after each call, so the same event $B_t$
covers all neighbour calls from update $t$.

For each node $j$, let $\mathcal{E}_j$ be the set of its uncached updates
through time $T$, and set $V_s=\{j:\mathcal{E}_j\ne\varnothing\}$.
Call node $j\in V_s$ open on the event
$O_j=\bigcup_{t\in\mathcal{E}_j}B_t$.
By \eqref{eq:dominating-events}, these events are conditionally
independent, with
\begin{equation}
\label{eq:node-open-probability}
\pi_j:=\operatorname{pr}(O_j\mid\mathcal{G}_s)
=1-\exp\left(-\sum_{t\in\mathcal{E}_j}\omega_t\right).
\end{equation}
Let $\mathcal{Q}_s\subseteq V_s$ be the nodes of the remaining requests
whose records are absent from $\mathcal{C}_s$.
Write $I_j$ for the indicator that at least one new record at node $j$
is resolved while completing these requests.
Any such node can be reached from $\mathcal{Q}_s$ along a path whose
vertices before $j$ are open.
Let $\Gamma_s(q,j)$ contain the paths from $q$ to $j$ with no repeated
vertex in the graph induced by $V_s$.
Conditional independence along each path and a union bound give
\begin{equation}
\label{eq:node-reachability-bound}
\operatorname{pr}(I_j=1\mid\mathcal{G}_s)\le h_j
:=\min\left\{1,\;
\sum_{q\in\mathcal{Q}_s}
\sum_{\gamma\in\Gamma_s(q,j)}
\prod_{v\in\gamma\setminus\{j\}}\pi_v
\right\}.
\end{equation}
The length-zero path at a queried node has weight one; a sum over
no paths is zero.
The product excludes the terminal node because resolving its record
incurs work before any further propagation is needed.
Theorem~\ref{thm:conditional-work} converts these reachability bounds
into work bounds, including range generation at unresolved neighbours.

\begin{theorem}
\label{thm:conditional-work}
Under the conditions of Theorem~\ref{thm:query-correctness} and the
preceding construction, let $W_s$ be all additional work
from $\mathcal{C}_s$ until the remaining requests and their statistical
outputs are completed, in the additive cost model of
\S\ref{sec:queries-work}.
Let $b_s,c_j,g_j$ be finite, nonnegative, $\mathcal{G}_s$-measurable
cost bounds, valid uniformly over all remaining inputs.
Here $b_s$ covers additional global preparation, query overhead and
output work; $c_j$ covers backend and management work charged when
new records at $j$ are resolved; and $g_j$ covers separately charged
proposal-input and range generation at $j$.
Each node bound covers the whole remaining block, including repeated
operations, and each operation is charged once.
Let $J_j$ indicate that the separately charged generation work occurs.
Let $\mathcal{D}_s$ be a $\mathcal{G}_s$-measurable directed relation
on $V_s$, where $(i,j)$ records a possible trigger from resolution at
$i$ to generation at $j$.
Assume
$J_j\le I_j+\sum_{i:(i,j)\in\mathcal{D}_s}I_i$, and the work satisfies
\begin{equation}
\label{eq:stage-work-accounting}
W_s\le b_s+\sum_{j\in V_s}c_jI_j+\sum_{j\in V_s}g_jJ_j.
\end{equation}
Set $h_j^+=\min\{1,h_j+\sum_{i:(i,j)\in\mathcal{D}_s}h_i\}$.
Then, almost surely,
\begin{equation}
\label{eq:conditional-work-bound}
\mathrm{E}(W_s\mid\mathcal{G}_s)
\le b_s+\sum_{j\in V_s}\bigl(c_jh_j+g_jh_j^+\bigr).
\end{equation}
\end{theorem}

First entries into unresolved nodes form a forest rooted at the requested
nodes. Each edge requires an open parent, which yields the path bound
\eqref{eq:node-reachability-bound}. Applying the generation trigger condition
and taking conditional expectations proves the result; see Supplementary
Material, \S\ref{supp:conditional-proof}.

\subsection{Finite histories on bounded-degree graphs}
\label{sec:bounded-degree-histories}

We now average the conditional work bound over independent proposal inputs.
Fix integers $\Delta\ge2$ and $L\ge1$.
Consider \eqref{eq:graph-posterior} with $a_j=1$, maximum degree at most
$\Delta$, and edge weights $0<w_{jk}\le w<\infty$.
Use independent standard Gaussian parameter proposals and a prescribed
schedule with at most $L$ updates per node.
Start from the initial-record cache $\mathcal{C}_0$, with
$|\theta_j^{(0)}|\le B$ for a fixed $B\ge0$.
The graph, initial records, schedule and ordered query set $Q$ are fixed
before the random inputs are generated.
Let $F_j$ contain the raw parameter and auxiliary proposal inputs at node
$j$, padded to $L$ updates.
Assume that these blocks are independent and identically distributed,
and independent of the acceptance thresholds.
Suppose that the graph admits $D$ colours such that distinct vertices at
graph distance at most two have different colours.
One may always take $D=\Delta^2+1$.
For independent standard Gaussian variables $Y_1,\ldots,Y_L$, let
$A=2(B+\max_{1\le\ell\le L}|Y_\ell|)$ and define
\[
q_D(w)=
\left(\mathrm{E}\left[\left\{
1-\exp\left(-wLA_0\sum_{i=1}^{\Delta}A_i\right)
\right\}^{D}\right]\right)^{1/D},
\]
where $A_0,\ldots,A_\Delta$ are independent copies of $A$.

\begin{corollary}
\label{cor:bounded-degree-work}
Under the conditions of Theorem~\ref{thm:query-correctness} and the
preceding construction, suppose that $(\Delta-1)q_D(w)<1$.
Assume that $C(F_j)$ bounds all work needed to generate and evaluate
the at most $L$ candidate records at node $j$, excluding work already
charged to $W_{\mathrm{setup}}$.
The same nonnegative function $C$ applies to every node and graph size,
and $c=\mathrm{E}\{C(F_j)\}<\infty$.
Use unit-cost exact arithmetic and indexed access, and amortized unit-cost
cache operations on fixed-size records.
Charge the preparation of the graph, data, initial records, random inputs
and event indices to $W_{\mathrm{setup}}$.
For the implementation of Algorithm~\ref{alg:query-execution} using
per-node event indices and cached ranges, the incremental work of resolving
$Q$ from $\mathcal{C}_0$ satisfies
\[
\mathrm{E}\{W_{\mathrm{inc}}(Q)\}\le\kappa |Q|,
\]
where $\kappa<\infty$ depends only on $\Delta,D,L,B,w,c$ and the fixed
operation costs.
The bound is uniform over the graph size $n$, graphs, schedules and
query sets satisfying these conditions.
\end{corollary}

The proof bounds each node's opening probability using its own input
block and those of its neighbours. A distance-two colouring makes these
bounds independent within each colour. H\"older's inequality and
self-avoiding path counts give the contraction condition; dropping the
factors involving an endpoint controls its candidate cost.
The argument is detailed in Supplementary Material, \S\ref{supp:colour}.
A finite rectangular four-neighbour grid has $\Delta=4$ and admits $D=5$:
assign colour $i+2j\pmod{5}$ to the vertex with integer coordinates $(i,j)$.
Since $q_D(w)\to0$ as $w\downarrow0$, the condition holds for sufficiently
small positive coupling strengths.
A directly checkable sufficient condition is
\[
4wL\Delta(\Delta-1)
\left\{B+(Lm_D)^{1/D}\right\}^{2}<1,
\]
where $m_D=\mathrm{E}(|Z|^D)$ for $Z\sim N(0,1)$.
This follows from $1-\exp(-x)\le x$ for $x\ge0$ and a bound on the
$D$th moment of the largest absolute proposal.
Evaluating $q_D(w)$ directly retains the exponential term and can certify
a wider coupling range.
For a fixed number of systematic scans, fixed $B,w$ and a common bound on
candidate generation and evaluation costs satisfying the corollary,
the same $\kappa$ applies as the grid's side lengths grow.

\subsection{Block execution on paths}
\label{sec:path-blocks}

We specialize Algorithm~\ref{alg:query-execution} to the path
$1,\ldots,n$, with $a_j=1$, common edge weight $0<w<\infty$ and
given initial records with $\theta_j^{(0)}=0$.
The reference schedule visits $n,n-1,\ldots,1$, performing $L\ge1$
consecutive updates at each node; the query is $Q=\{(1,nL)\}$.
Let $Y_{j\ell}$, $j=1,\ldots,n$ and $\ell=1,\ldots,L$, be
independent standard Gaussian parameter candidates.
An addressable source supplies independent raw proposal inputs and
thresholds with common laws across nodes. Addresses specify node,
local update and input type; repeated access returns the same input.
Input blocks are cached when first needed, and candidates are evaluated
as their updates are resolved.
During block $j$, the left neighbour $j-1$, when present, has its
initial parameter value zero at the reference times of these updates.
The right neighbour $j+1$, when present, has its block-end value.
Until that block is resolved, its parameter is enclosed by the interval
spanned by $0,Y_{j+1,1},\ldots,Y_{j+1,L}$.
An undecided comparison triggers resolution of block $j+1$, then resumes
with the same old record, candidate and threshold.

The complete block procedure is given in Algorithm~\ref{alg:path-block-execution}
 in the Supplementary Material.

Let $F,F'$ be independent copies of a node's raw proposal-input block.
For the parameter coordinates $Y_1,\ldots,Y_L$ of $F$, set $Y_0=0$ and
define
\[
r(F)=\max_{0\le\ell\le L}Y_\ell-\min_{0\le\ell\le L}Y_\ell,
\qquad
s(F)=\sum_{\ell=1}^{L}\max_{0\le h<\ell}|Y_\ell-Y_h|.
\]
Here $r(F)$ is the block's range width, and $s(F)$ bounds the sum of
distances between its candidates and their possible old parameters.
For the $\ell$th update of block $j<n$, choose $\omega_t$ in
\eqref{eq:canonical-band-width} to be
$w r(F_{j+1})\max_{0\le h<\ell}|Y_{j\ell}-Y_{jh}|$,
where $t=(n-j)L+\ell$ and $Y_{j0}=0$.
These bounds sum to $w s(F_j)r(F_{j+1})$; at the terminal node, take $\omega_t=0$.
Conditional on all raw input blocks, \eqref{eq:node-open-probability} bounds the canonical opening probability of each block $j<n$ by $p(F_j,F_{j+1})$, where
\[
p(F,F')=1-\exp\{-w s(F)r(F')\},
\qquad
\rho=\bigl[\mathrm{E}\{p(F,F')^2\}\bigr]^{1/2}.
\]
For every fixed finite $w$ and $L$, these ranges are finite almost surely, so $0\le\rho<1$.
Adjacent factors share one input block; the second moment in $\rho$ controls this dependence.

\begin{theorem}
\label{thm:path-block-work}
Consider Algorithm~\ref{alg:path-block-execution} under the preceding
path specification and the conditions of
Theorem~\ref{thm:query-correctness}.
Use exact real arithmetic, unit-cost access to the addressable input
source and fixed-size records.
The supplementary accounting model assigns one logical management unit
to each raw input draw, range visit, block lookup, threshold read,
candidate dispatch, certificate comparison, record commit and listed
control operation.
Let $C(F_j)$ bound all remaining work of constructing and evaluating
the $L$ candidates when block $j$ is resolved, uniformly over its
possible old records.
Assume the same nonnegative function $C$ applies at every node, with
$c=\mathrm{E}\{C(F)\}<\infty$, and define
\[
\delta=\mathrm{E}\{C(F')p(F,F')\},
\qquad
\varepsilon=\mathrm{E}\{C(F')p(F,F')^2\}.
\]
Let $N$ be the number of resolved blocks.
Then, for every integer $h\ge0$,
\begin{equation}
\label{eq:path-block-tail}
\operatorname{pr}(N\ge h+1)\le\rho^h,
\qquad
\mathrm{E}(N)\le\frac{1}{1-\rho},
\end{equation}
and the incremental work satisfies
\begin{equation}
\label{eq:path-total-work}
\mathrm{E}\{W_{\mathrm{inc}}(Q)\}
\le c+\delta+\frac{\rho(c\varepsilon)^{1/2}}{1-\rho}
+24L\left(1+\frac{1}{1-\rho}\right).
\end{equation}
For fixed $w,L$, input law, common envelope $C$ and operation costs,
this bound is uniform in $n$.
Preparation and output work retain their separate terms in
\eqref{eq:query-work}.
\end{theorem}

The proof groups propagation factors by odd and even edges, whose inputs
are independent within each group. Weighted Cauchy--Schwarz retains the
dependence between the final edge and its candidate cost. At most $N+1$
input blocks are generated, including any unresolved boundary block, and
management takes at most $24L(N+1)$ logical units.
The full argument and operation counts are in Supplementary Material,
\S\ref{supp:path-proof}.

\subsection{Comparison with pointwise ancestor evaluation}
\label{sec:pointwise-comparison}

We compare certificate execution with a pointwise ancestor evaluator
$B_{\mathrm{point}}$.
Both evaluators know the model, graph, schedule and ordered query set $Q$.
They start from the same prepared initial-record cache $\mathcal{C}_0$
and use the same proposal inputs and acceptance thresholds, under the
same input-access and cost conventions.
For each requested uncached update $t$, $B_{\mathrm{point}}$ first
resolves its own predecessor, then generates and evaluates the
candidate once.
It shares a cache of complete records across requests and retains the
candidate and threshold across recursive calls.
A zero estimate is rejected immediately.
When the proposed parameter equals the old parameter, the decision
uses the two local records and $U_t$.
For every other candidate with a positive likelihood estimate at node $j=j_t$, it first resolves
all neighbour records $R_k^{(p_k(t))}$, $k\in\mathcal{N}_j$,
reusing cached records and processing uncached neighbours in a fixed order.
It then evaluates \eqref{eq:reference-acceptance} and compares the
acceptance probability with $U_t$.
Accepted candidates and retained old records are cached as in
Algorithm~\ref{alg:query-execution}.
The following corollary compares the incremental work of the two
execution rules for the same indexed collection $\mathcal{R}_Q$.

\begin{corollary}
\label{cor:pointwise-separation}
Suppose that the candidate likelihood estimate is strictly positive
almost surely at every scheduled update.
Consider either of the following families, with the input and cost
assumptions of the cited result.
On the paths of Theorem~\ref{thm:path-block-work}, use
Algorithm~\ref{alg:path-block-execution} and $Q=\{(1,nL)\}$.
On rectangular four-neighbour grids satisfying
Corollary~\ref{cor:bounded-degree-work}, number the $n$ vertices row by
row from top left to bottom right, perform $L$ scans in the order
$1,\ldots,n$, and use
Algorithm~\ref{alg:query-execution} with $Q=\{(n,nL)\}$.
Implement $B_{\mathrm{point}}$ with indexed record access and acceptance
ratios evaluated from the local factors.
Let $N_{\mathrm{call}}^{\mathrm{point}}$ count its candidate likelihood
evaluations, and let $W_{\mathrm{inc}}^{\mathrm{point}}(Q)$ and
$W_{\mathrm{inc}}^{\mathrm{cert}}(Q)$ denote the respective incremental
work of the pointwise and certificate evaluators.
Within each family, keep $L$, the input law, the common envelope $C$
and all other parameters in the respective work bound fixed as $n$ varies.
Then
\begin{equation}
\label{eq:pointwise-separation}
\begin{aligned}
N_{\mathrm{call}}^{\mathrm{point}}&=nL
    \quad\text{almost surely},\\
\mathrm{E}\{W_{\mathrm{inc}}^{\mathrm{point}}(Q)\}&=\Theta(nL),
\qquad
\sup_n\mathrm{E}\{W_{\mathrm{inc}}^{\mathrm{cert}}(Q)\}<\infty.
\end{aligned}
\end{equation}
Both evaluators return the reference collection $\mathcal{R}_Q$.
\end{corollary}

On paths, the pointwise rule propagates through successive right-hand
blocks. On grids, the final query reaches the last-scan updates through
west and north neighbours. Resolving own predecessors then includes all
$nL$ candidates. The common cost envelope and the certificate bounds give
\eqref{eq:pointwise-separation}; Supplementary Material,
\S\ref{supp:separation-proof}, provides the full proof.

Table~\ref{tab:work-bounds} in the Supplementary Material summarizes
the conditions and work bounds.

\section{Numerical experiments}
\label{sec:numerical-experiments}

\subsection{Block resolution and computational work}
\label{sec:block-resolution-work}
The experiments examine how historical resolution, execution strategy and
query workload affect computational work.

\begin{figure}[tbp]
\centering
\setlength{\figwidth}{\linewidth}
\includegraphics[width=\figwidth]{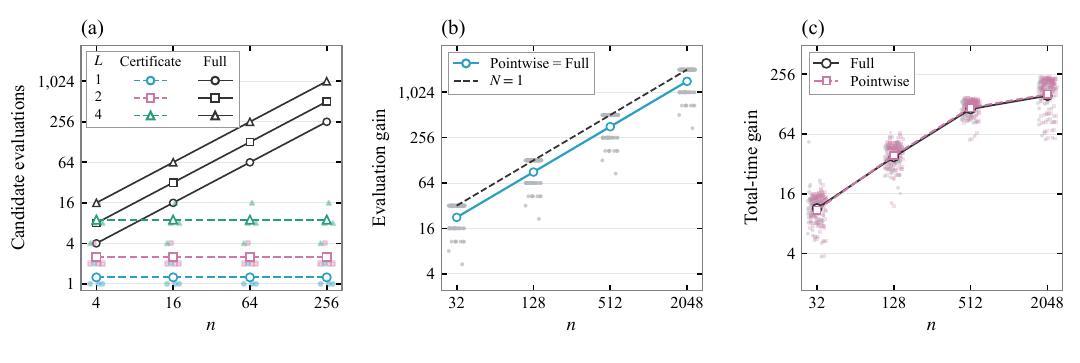}
\caption{Work and time gains for a local query on paths.
The query requests the complete record at node 1 at reference time $nL$ on a path with $n$ nodes and $L$ updates per node.
Panel (a) uses an estimate that can be zero, $w=0.2$, $L\in\{1,2,4\}$ and the same four fixed input streams.
Panels (b) and (c) use a strictly positive estimate, $w=0.2$, $L=4$ and 100 common input streams.
Panel (a) shows candidate counts; (b) shows evaluation gains; (c) shows total-time gains, including preparation.
Gains are baseline costs divided by certificate costs.
Pointwise and full candidate counts coincide in (b).
Small markers show individual streams, displaced horizontally to separate overlapping points.
In (a), open markers joined by lines show mean counts.
In (b) and (c), they show ratios of mean costs.
The dashed line in (b) marks a gain of $n$, attained when only one block is resolved ($N=1$).
Both axes are logarithmic.}
\label{fig:block-work}
\figurealt{Panel (a) shows stable certificate evaluation counts and increasing full execution counts as paths grow. Panel (b) shows increasing evaluation gains over pointwise and full execution. Panel (c) shows more slowly increasing total time gains, with variation across input streams.}
\end{figure}

We begin with the path model and block schedule in \S\ref{sec:path-blocks},
with $a_j=1$, common edge weight $w$, and independent standard Gaussian
parameter proposals.
At each node, set
\[
\ell_j(\theta)=\exp(-\theta^2/2),
\qquad
\widehat{\ell}_j(\theta,u)
=2\,\mathbf{1}\{u<1/2\}\exp(-\theta^2/2),
\]
where $u$ is uniformly distributed on $(0,1)$.
This gives an unbiased estimate with probability $1/2$ of being zero.
All initial records are $R_j^{(0)}=(0,1/4,2)$.
We compare Algorithm~\ref{alg:path-block-execution} with full execution
for $Q=\{(1,nL)\}$, using the same parameter candidates, auxiliary
inputs and acceptance thresholds.
We take $n\in\{4,16,64,256\}$, $L\in\{1,2,4\}$ and
$w\in\{0.01,0.2,1.5\}$.
We use the same four fixed input streams for all configurations.
Inputs at common node and update addresses are identical.
Figure~\ref{fig:block-work}(a) compares candidate counts at $w=0.2$ for three values of $L$.
Supplementary Material, \S\ref{supp:counts}, reports the results for all three edge weights.

Algorithm~\ref{alg:path-block-execution} evaluates $LN$ candidates
for $N$ resolved blocks; full execution evaluates $nL$.
Across the 144 fixed configurations, $N$ ranges from 1 to 4.
For each fixed $L$, $w$ and input stream, $N$ is unchanged across
the four path lengths.
The returned complete record agrees with full execution in every
configuration.

Management work ranges from 12 to 176 logical units, within the bound
$24L(N+1)$ of Theorem~\ref{thm:path-block-work}.
This count includes input and range generation for any unresolved boundary block.
Likelihood evaluation, preparation and statistical output costs retain
their separate terms in \eqref{eq:query-work}.

In a second experiment, we use the strictly positive estimate $\widehat{\ell}_j(\theta,u)=\exp(-\theta^2/2+u-1/2)$, where $u\sim N(0,1)$.
We fix $L=4$ and $w=0.2$, and use 100 common input streams for the same query on paths with $n\in\{32,128,512,2048\}$.

At every tested path length, certificate execution resolves a mean of 1.43 blocks and at most six blocks.
Its mean candidate count is 5.72, compared with $4n$ for the specified pointwise evaluator and full execution.
The three executors return the same complete query record in each of the 400 matched comparisons.
Figure~\ref{fig:block-work}(b) shows the resulting gains in candidate evaluations as $n$ increases.
Panel (c) shows the gains in total elapsed time, including preparation of all $n$ initial records.
At $n=2048$, mean pointwise and full times are 160.7 and 155.7 times the mean certificate time, respectively.
In this implementation, preparation accounts for 5.2\% of mean certificate time at $n=32$ and 73.3\% at $n=2048$.
Supplementary Material, \S\ref{supp:counts}, gives the operation counts and the timing breakdown.
We next examine these costs with particle likelihood estimators.

\subsection{Execution cost for the same queries}
\label{sec:same-query-cost}

We compare cached pointwise and certificate execution for two models:
a Gaussian latent process with Poisson observations, and a gene expression
process with switching activity and binomial capture.
Both use the Gaussian graph prior in \eqref{eq:graph-posterior} on a path
with 128 nodes, $a_j=1$ and $w_{jk}=0.8$.
Particle filters provide likelihood estimates \citep{AndrieuDoucetHolenstein2010}.
Two simulated datasets per model give six configurations of data, particle estimator and parameter proposal.
For the gene expression model, the two particle estimators also use different parameter proposals.
Full model and computational settings are given in Supplementary Material, \S\ref{supp:experiments}.
Each run returns the parameter histories of nodes 1 and 2 at every
reference time.
We use five paired runs per configuration, with common initial complete
records, update schedules, proposal inputs and acceptance thresholds
within each pair.
The pointwise evaluator of \S\ref{sec:pointwise-comparison} and Algorithm~\ref{alg:query-execution} both cache complete records and reuse evaluated candidates.
Pointwise and certificate execution use four worker processes for initialization and evaluate subsequent candidates in the controlling process.
Elapsed times include input preparation, initialization, construction of random inputs and indices, range maintenance, query evaluation, statistical summaries, and process startup and shutdown.

\begin{figure}[tbp]
\centering
\setlength{\figwidth}{\linewidth}
\includegraphics[width=\figwidth]{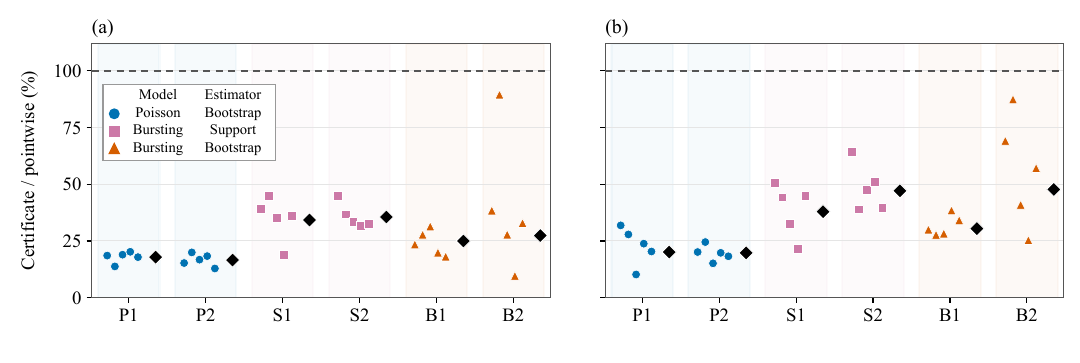}
\caption{Candidate evaluations and elapsed time for the same queries.
Both executors return the parameter histories of nodes 1 and 2 at every reference time on a path with 128 nodes.
Panels (a) and (b) show candidate likelihood evaluations and total elapsed time, respectively.
Coloured points show certificate costs as percentages of pointwise costs in five paired runs per configuration.
Black diamonds to the right of each group show ratios of mean costs, expressed as percentages.
The dashed lines mark equal cost at 100\%.
P1 and P2 use two datasets from the Poisson observation model.
For the bursting gene expression model, S1 and B1 share one dataset, and S2 and B2 share the other.
The \textit{support} filter conditions on the observations; S1 and S2 use parameter proposals independent of the current value.
All bootstrap configurations use random walk proposals.
Within each configuration, the five paired runs appear from left to right in the same order in both panels.}
\label{fig:same-query-cost}
\figurealt{Panel (a) shows fewer candidate evaluations for certificate execution in every paired run across six configurations. Panel (b) shows shorter elapsed times in every pair. Black diamonds show smaller percentage reductions in mean time than in mean evaluation count.}
\end{figure}

The returned parameter histories agree in all 30 paired runs.
Figure~\ref{fig:same-query-cost} shows candidate evaluation counts and total elapsed times for certificate execution relative to pointwise execution.
Across the six configurations, certificate execution uses 16.6--35.6\% of the mean number of candidate evaluations required by pointwise execution.
Each executor evaluates one candidate per resolved historical update.
Its mean elapsed time is 19.7--47.7\% of the pointwise time, and it is faster in every pair.
The ratio of mean times exceeds the ratio of mean candidate counts in every configuration.
Absolute timings and their dispersion are given in Supplementary Material, \S\ref{supp:timings}.

We examine the contribution of cached ranges using 128 saved trajectories (Supplementary Material, \S\ref{supp:experiments}).
The executor settles 163,164 decisions while some required neighbour records remain unresolved.
At each of these comparisons, we keep the old record, candidate record and acceptance threshold fixed and recompute the bounds using ranges propagated from the initial states.
The replacement ranges leave 101,784 of these decisions (62.4\%) undecided.
We next vary the set of queried nodes to examine its effect on
computational work.

\subsection{Query scope and execution cost}
\label{sec:query-scope-cost}

We vary the number of queried nodes in P1 and P2 and compare certificate, pointwise and Picard execution.
The Picard executor adapts the exact Online Picard algorithm \citep{GrazziZanella2026} to the complete pseudo-marginal state.
It evaluates candidates in parallel within a moving window and advances the resolved prefix.
Our implementation uses four worker processes, a window of 16 updates and a cache of evaluated candidates.
The model settings and timing convention are those of \S\ref{sec:same-query-cost}.

Each task requests the parameter histories of nodes 1 through $m$ at every reference time, with $m\in\{2,8,32,64,128\}$.
We use five random-input streams per dataset and the same streams at all five query scopes.
Within each stream, all three executors share the initial complete records, update schedule and fixed random inputs.
Each execution starts with a fresh cache and pays for initialization.
The returned parameter histories agree across executors in all 50 combinations of dataset, input stream and query scope.

\begin{figure}[tbp]
\centering
\setlength{\figwidth}{\linewidth}
\includegraphics[width=\figwidth]{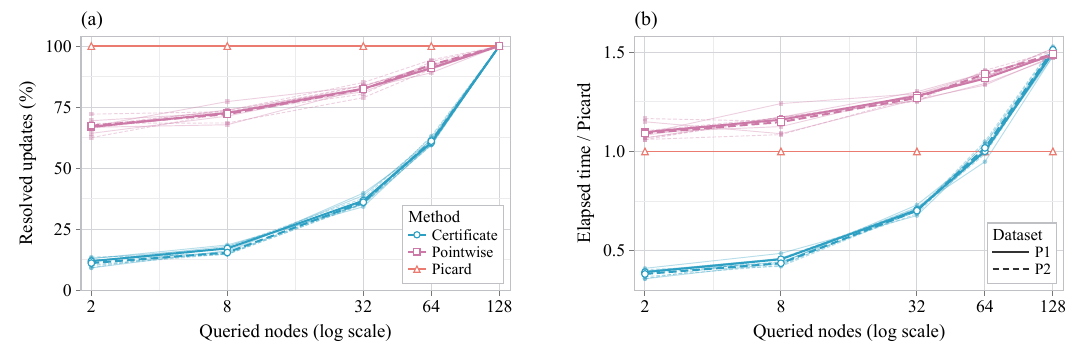}
\caption{Query scope, resolved history and execution cost.
Each task requests the parameter histories of nodes 1 through $m$ at every reference time in P1 or P2.
Panel (a) shows the percentage of scheduled updates resolved; panel (b) shows total elapsed time relative to Picard.
Five input streams per dataset are shared across methods and query scopes.
Thin lines and small points show individual certificate and pointwise runs.
Thicker lines with open markers show mean percentages in (a) and ratios of mean times in (b).
The method and dataset legends apply to both panels.
The Picard curves coincide at 100\% in (a) and at 1 in (b).
The horizontal axes use a logarithmic scale; the vertical axes are linear.}
\label{fig:query-scope-cost}
\figurealt{Panel (a) shows certificate and pointwise execution resolving increasing fractions of history as queries expand, eventually reaching the full history. Panel (b) shows certificate execution changing from faster than Picard for small queries to slower for full queries in both datasets.}
\end{figure}

Figure~\ref{fig:query-scope-cost} shows how historical work and elapsed time relative to Picard change with query scope.
For two-node queries, pointwise execution resolves about 67\% of the schedule on average in each dataset, compared with 11--12\% for certificate execution.
Both executors resolve the full schedule when all 128 node histories are requested.
Picard resolves the full schedule at every query scope.

Certificate execution is faster than Picard in every matched run for 2, 8 and 32 queried nodes.
At 64 nodes, certificate execution resolves about 61\% of the schedule on average in each dataset.
Its mean elapsed time is then close to Picard's in both datasets.
For full queries, mean certificate times are 1.497 and 1.512 times the corresponding mean Picard times in P1 and P2, respectively.

In all six configurations of \S\ref{sec:same-query-cost}, certificate execution takes less time on average than Picard for two-node queries (five paired runs per configuration).
Picard is faster for full queries in the single paired run per configuration.
The two executors return identical parameter histories in all 36 pairs.

In P1 and P2, increasing the Picard window from 16 to 64 updates reduces elapsed time by 16.7--24.5\% across the two-node and full queries (four workers, one run per setting).
Supplementary Material, \S\ref{supp:timings}, gives the individual timings.

We next examine historical work and elapsed time for the statistical
queries in the applications.

\section{Applications}
\label{sec:applications}

\subsection{Regional inference in positron emission tomography}
\label{sec:pet-application}

We study regional quantification in positron emission tomography (PET)
using one baseline $[^{11}\mathrm{C}]\mathrm{MC1}$ scan from OpenNeuro
dataset ds004869 \citep{YanEtAl2025MC1,YanEtAl2026MC1Data}.
The analysis uses 27 frames covering the first 90\,min, together with
the supplied plasma input and whole-blood curves.
At each voxel, the total distribution volume $V_T$ is the equilibrium
ratio of tissue to plasma tracer concentration under a constant plasma
input \citep{ZhouAstonJohansen2013}.
Our query is the average of $V_T$ over the 1311 voxels in the predefined
left insula region.
This region lies within an analysis domain of 41\,500 voxels in the
left hemisphere.
We examine how uncertainty about the number of tissue compartments
and spatial dependence between model labels affect inference for this
regional average.

We use one- and two-tissue compartment models with Gaussian observation errors, conditioning on the supplied input curves and observation weights.
Kinetic parameters have independent log-uniform priors on bounded intervals, and noise precisions have gamma priors.
The binary model labels have equal base probabilities and a Potts prior that favours agreement between face-adjacent voxels, with coupling parameter $0.25$.
Inference uses a Markov chain on the model labels, with the kinetic parameters and noise precisions integrated out.
Analytical integration over each precision and deterministic positive quadrature over the kinetic parameters give the model evidences.
At each update, the reference chain selects a voxel uniformly and proposes either label with probability $1/2$.
Certificate execution bounds the Potts contribution from unresolved neighbours and resolves their historical labels only when the acceptance test requires them.

For each voxel and model, we group the $V_T$ quadrature masses into fixed logarithmic bins and place each bin's mass at its geometric midpoint.
These tables and the spatial label posterior define the numerical posterior used below.
At each prescribed query time, we resolve the regional label vector, draw independently from the corresponding conditional $V_T$ tables and average these voxel values.
The resulting regional draws incorporate spatial dependence between labels and conditional variation in $V_T$.
We estimate the posterior mean by averaging the conditional regional means and estimate interval endpoints from quantiles of the regional draws.
Supplementary Material, \S\ref{supp:pet}, gives the model details and numerical rules.
It also proves agreement of the requested labels and regional draws with full execution from the same initial labels and random inputs.

Four chains of 512 sweeps, with the first 64 discarded in each chain,
give a spatial posterior mean of
$2.05372\,\mathrm{mL}\,\mathrm{cm}^{-3}$ and a 95\% equal-tailed
interval of $[2.04246,2.06468]$.
Fixing all voxels to the two-compartment model gives a mean of
2.05449, whereas setting the spatial coupling to zero and averaging
over model labels gives 2.05323.
The largest difference between these means is 5.7\% of the spatial
posterior interval width.
Under the spatial model, variation in $V_T$ conditional on the labels
accounts for 99.7\% of the estimated posterior variance of the
regional average.
The estimated Monte Carlo standard error of the regional posterior mean is $1.23\times10^{-5}\,\mathrm{mL}\,\mathrm{cm}^{-3}$.
With the model evidences and sampled labels fixed, using unbinned conditional moments lowers the mean estimate by about $1.5\%$ of this standard error.
Supplementary Material, \S\ref{supp:diagnostics}, gives the rank-based chain diagnostics and interval endpoint error estimates \citep{VehtariEtAl2021}.
Figure~\ref{supp:pet-quadrature} in \S\ref{supp:pet} compares two quadrature rules for the conditional $V_T$ distributions and model evidences at 12 prespecified regional voxels.

\begin{figure}[tbp]
\centering
\setlength{\figwidth}{\linewidth}
\includegraphics[width=\figwidth]{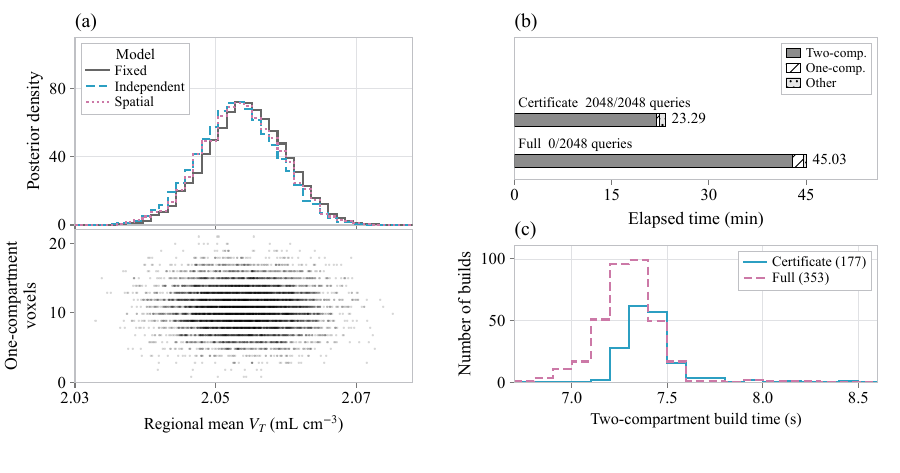}
\caption{Regional PET inference and its computational cost.
PET denotes positron emission tomography.
(a) Upper: posterior density histograms of regional mean $V_T$ under fixed
two-compartment modelling, independent model averaging and spatial
model averaging, using 8192, 8192 and 7168 draws, respectively.
Lower: the 7168 spatial joint draws of regional mean $V_T$ and the number
of voxels assigned to the one-compartment model.
(b) Observed elapsed time, partitioned into completed two-compartment
evidence builds, one-compartment evidence builds and all remaining
elapsed time (Other).
Both executors start from the same cache and target the first chain's 2048 regional queries.
Certificate execution completes 2048/2048 queries; full execution completes 0/2048 before reaching its time budget.
Construction of the shared cache and integration basis is excluded.
(c) Histograms of the durations of 177 and 353 completed two-compartment
evidence builds for certificate and full execution, respectively.
Build times include numerical integration and database writes.
Histograms use common bins within each comparison.}
\label{fig:pet-regional-cost}
\figurealt{Panel (a) shows similar regional posterior distributions under three models. Below, joint draws relate the regional mean to the number of voxels assigned to one compartment. Panel (b) shows certificate execution completing all queries, while full execution reaches its time budget with none completed. Panel (c) shows overlapping construction time distributions, with fewer evidence builds under certificate execution.}
\end{figure}

We compare the two executors on the first chain, using identical
initial states, random inputs and 2048 prescribed regional queries.
Both start from the same cache of 2942 voxel--model evidence values
and regional conditional distributions, and use the same integration
routines.
Certificate execution completes all queries in 23.29\,min.
Full execution stops at its budget after 45.03\,min, having completed
725 updates; the first query is scheduled after update 10,375.
Certificate execution builds 177 new two-compartment evidence values for the regional queries.
Full execution builds 353 as it advances through the graph schedule before the first regional query.
Figure~\ref{fig:pet-regional-cost}(a) shows the regional posterior distributions and how regional $V_T$ varies with model composition.
Completed two-compartment builds account for over 90\% of the observed time in each run (panel b).
Mean build times are 7.4\,s for certificate execution and 7.3\,s for full execution (panel c).
Elapsed time runs from process launch to exit, with shared preparation costs reported separately in Supplementary Material, \S\ref{supp:pet-cost}.

\subsection{Comparing covariance scales in the Landmine data}
\label{sec:landmine-application}

We compare the covariance length scales of two sites in the Landmine radar data \citep{XueEtAl2007}.
Each observation comprises nine features and a binary label indicating a landmine or clutter.
We use 200 observations from each of sites 1--15, which represent highly foliated regions.
Features are standardized using the pooled training observations.
At site $j$, we use a probit Gaussian process model whose squared exponential covariance component has length scale $\exp(z_j)$ and signal standard deviation $\exp(\gamma_j)$.
We assign conditionally independent priors $z_j\mid\mu\sim N(\mu,0.75^2)$, with $\mu\sim N(\log3,0.75^2)$, where $\mu$ is the shared mean.
The log amplitudes $\gamma_j$ have independent standard normal priors.
Integrating out $\mu$ gives a Gaussian prior on a complete graph for the centred log length scales $z_j-\log3$.
We query the joint state $(z_1,z_2)$ and summarize the posterior of $d=z_1-z_2$ by its mean, 95\% equal-tailed interval and probability of being positive.
Positive values of $d$ indicate a longer covariance length scale at site 1 in the standardized feature space.

At each site, expectation propagation supplies a Gaussian importance distribution for the latent values.
We average the complete importance weights of 32 independent draws to obtain a nonnegative, unbiased marginal likelihood estimator, following \citet{FilipponeGirolami2014}.
Each parameter update jointly proposes $(z_j,\gamma_j)$ from a fixed mixture distribution independent of the current parameter value.
Only the log length scales enter the prior terms that couple sites.
Certificate execution therefore bounds neighbouring log length scales and evaluates the amplitude prior and full mixture density at the old and proposed parameter pairs.
On rejection, it retains the complete old parameter and auxiliary record.
Supplementary Material, \S\ref{supp:landmine}, gives the centred prior, acceptance bounds and correctness argument.

For comparison, we use ancillary augmentation (AA), with covariance updates conditional on whitened Gaussian variables \citep{MurrayAdams2010}.
The latent process and its Gaussian intercept are updated separately by elliptical slice sampling \citep{MurrayAdamsMacKay2010}.
At each site, ten such pairs of updates precede a joint Metropolis--Hastings update of $(z_j,\gamma_j)$.
Both samplers target the same marginal posterior for the covariance parameters, and we run four chains of each.
Preprocessing, covariance details and sampling schedules are specified in Supplementary Material, \S\ref{supp:landmine}; diagnostics follow \S\ref{supp:diagnostics} of that supplement.

\begin{figure}[tbp]
\centering
\setlength{\figwidth}{\linewidth}
\includegraphics[width=\figwidth]{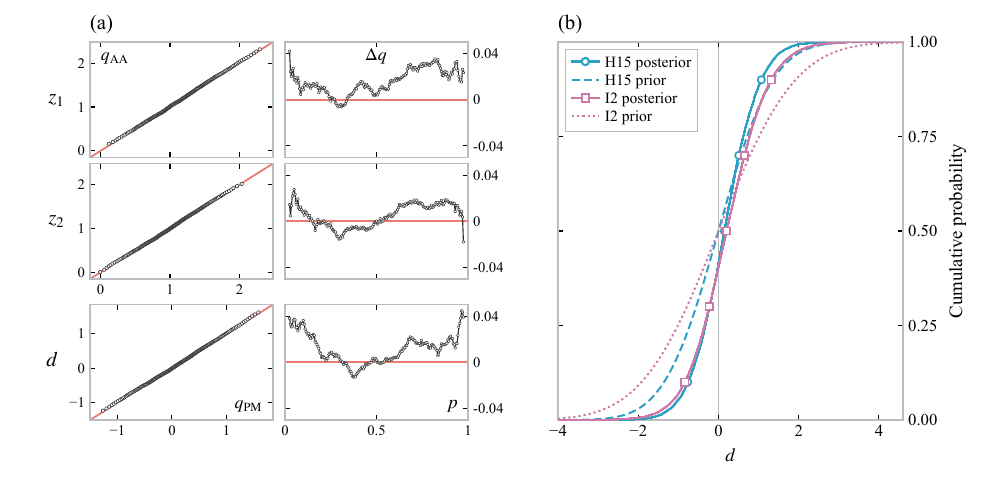}
\caption{Posterior distribution comparisons for the Landmine data.
(a) Comparisons of pseudo-marginal (PM) and ancillary augmentation (AA) samples under the shared 15-site model (H15).
Rows correspond to $z_1$, $z_2$ and $d=z_1-z_2$, from top to bottom.
The left column plots AA quantiles $q_{\mathrm{AA}}(p)$ against PM quantiles $q_{\mathrm{PM}}(p)$ at probability $p$.
The right column plots $\Delta q=q_{\mathrm{AA}}(p)-q_{\mathrm{PM}}(p)$ against $p$.
Both columns use 191 equally spaced values of $p$ from 0.025 to 0.975.
Red lines mark equality in the left column and zero difference in the right.
(b) Prior and posterior cumulative distribution functions of $d$ for H15 (blue) and the independent two-site model (I2, pink).
Solid curves show empirical posterior distributions from PM samples for both models.
Dashed and dotted curves show the H15 and I2 priors, respectively.
Circles and squares mark the H15 and I2 posterior quantiles at probability levels 0.1, 0.3, 0.5, 0.7 and 0.9.
The grey vertical line marks $d=0$.
H15 uses 24,576 PM and 294,912 AA draws; I2 uses 24,576 PM draws.
All samples are pooled over four chains after warm-up.}
\label{fig:landmine-posterior}
\figurealt{Panel (a) shows close agreement between pseudo-marginal and ancillary augmentation quantiles for both log length scales and their difference. Adjacent plots reveal small deviations across probabilities. Panel (b) compares prior and posterior distributions. The shared model has a more concentrated posterior for the difference than the independent model. Both posteriors retain substantial probability on either side of zero.}
\end{figure}

Table~\ref{tab:landmine-summary} compares pseudo-marginal (PM) and AA estimates of the mean of $d$, its two interval endpoints and $\operatorname{pr}(d>0)$ under the shared model.
Each difference is less than 1.5 combined estimated Monte Carlo standard errors in absolute value.
Figure~\ref{fig:landmine-posterior}(a) compares PM and AA quantiles of $z_1$, $z_2$ and $d$ and shows their signed differences.

We also fit sites 1 and 2 independently using the same observations, likelihoods and individual marginal priors for these sites.
Both models assign about 60\% posterior probability to a longer length scale at site 1 than at site 2.
The 95\% posterior interval for $d$ from PM draws is 16.7\% narrower under the shared model.
The shared model uses data from 13 additional sites and links the two log length scales through a common mean.
This link halves the prior variance of $d$ from 2.25 to 1.125.
Figure~\ref{fig:landmine-posterior}(b) shows the prior and posterior cumulative distribution functions under both models.

\begin{table}[tbp]
\setlength{\tabcolsep}{5pt}
\renewcommand{\arraystretch}{1.08}
\tbl{Posterior summaries of the Landmine site contrast}{%
\begin{tabular}{@{}llccc@{}}
Model & Sampler & Mean & 95\% interval & $\operatorname{pr}(d>0)$ \\[5pt]
Shared (H15)
  & PM
  & $0.152\ (0.009)$
  & $[-1.265,1.581]$
  & $0.587\ (0.005)$ \\
  & AA
  & $0.165\ (0.015)$
  & $[-1.227,1.620]$
  & $0.584\ (0.008)$ \\[3pt]
Independent (I2)
  & PM
  & $0.223\ (0.010)$
  & $[-1.441,1.973]$
  & $0.596\ (0.006)$
\end{tabular}}
\label{tab:landmine-summary}
\begin{tabnote}
The contrast $d=z_1-z_2$ compares the log length scales of sites 1 and 2.
H15 denotes the shared 15-site model and I2 the independent two-site model.
PM, pseudo-marginal; AA, ancillary augmentation.
Parentheses contain estimated Monte Carlo standard errors.
Intervals are equal-tailed posterior intervals.
Interval-endpoint errors and chain diagnostics are provided in
Supplementary Material, \S\ref{supp:diagnostics}.
\tablealt{In the shared model, both samplers give a posterior probability near 0.59 that the site contrast is positive. The 95\% intervals span zero in both models.}
\end{tabnote}
\end{table}

We query the two sites jointly after every 15 random-scan updates,
at 8192 times in each chain.
Across the four chains, certificate execution evaluates 491306 of
the 491520 scheduled candidates, leaving 214 (0.044\%) unevaluated.
Certificates settle 90.9\% of the evaluated updates while at least
one required neighbour record is unresolved.
Each update requires its own preceding PM record, so the resolved
history at each site forms a prefix of its update schedule.
The integrated shared prior also couples each site to the other 14.
Later recursion or queries can require records bypassed by an earlier
certificate, extending these prefixes (Supplementary Material, Fig.~\ref{supp:landmine-history}).
The computational workload depends on their accumulated extent over
the whole query sequence.

\section{Discussion}
\label{sec:discussion}

Decision certificates allow unresolved historical parameters to enter
acceptance calculations through ranges.
With the complete old and candidate records available, these ranges can
determine the decision for a fixed threshold.
Theorem~\ref{thm:query-correctness} establishes that this construction
preserves the complete queried records of the reference process.
In the path and grid families of Corollary~\ref{cor:pointwise-separation},
certificate execution and the specified pointwise ancestor evaluator
have different orders of expected incremental work.
Here the graph grows with per-node history length and all other
parameters in the work bounds fixed.
This separation holds for the same query, reference schedule and random
inputs, showing how the representation of unresolved neighbouring histories
can change the work required to recover a prescribed output.

\citet{AnandJerrum2022} sample configurations on finite windows from the Gibbs measure of a potentially infinite spin system.
Under exponential strong spatial mixing and subexponential graph growth, their expected running time is linear in the window size.
For finite reference trajectories, our graph and path bounds allow random candidate costs to depend on the proposal inputs governing propagation.
These bounds assume a common integrable cost envelope across nodes and graph sizes.

Certificate execution reuses cached records across a sequence of queries.
Later requests can also require histories bypassed by earlier certificates, as in Landmine.
Preparation costs and parallel evaluation also affect the time gains from local execution.
For larger paths, preparation accounts for a greater share of certificate time in the path experiment (\S\ref{sec:block-resolution-work}).
Broader queries shift the timing advantage to Picard in the Poisson experiments (\S\ref{sec:query-scope-cost}).
Total elapsed time for the same requested outputs captures these effects together.

The correctness results assume exact arithmetic.
The Landmine implementation uses outward-rounded bounds and stops if a fully resolved comparison remains undecided (Supplementary Material, \S\ref{supp:landmine}).

The conditional bound in Theorem~\ref{thm:conditional-work} provides a starting point for analysing cumulative work as node histories lengthen.
A sharper analysis would relate changes in cached ranges to the work required by later requests.
Such an analysis must track revealed thresholds and retained auxiliary records across requests and charge each operation once.
A complementary direction is to extend the addressable inputs of Algorithm~\ref{alg:path-block-execution} to broader classes of prescribed schedules.
This extension requires access to historical predecessors and valid ranges without generating the full random input sequence.
Combining this access with output routines driven by explicit requests would organize input generation, indexing and output work around the records needed for the statistical task.

\section*{Funding}
This research received no funding.

\section*{Conflict of interest}
The author declares no conflict of interest.

\section*{Data availability}
The tomography data are available from OpenNeuro as dataset ds004869,
version 1.4.1 \citep{YanEtAl2026MC1Data}.
The Landmine data are described by \citet{XueEtAl2007}.
The public mirror used here is specified in Supplementary Material,
\S\ref{supp:landmine}.
The replication package has the reserved DOI \url{https://doi.org/10.5281/zenodo.24506399}.

\section*{Supplementary material}
The Supplementary Material contains proofs, the block execution
procedure and operation counts, model and computational specifications,
individual timing records, and application diagnostics.

\bibliographystyle{biometrika}
\bibliography{paper-ref}
\clearpage
\setcounter{section}{0}
\setcounter{equation}{0}
\setcounter{figure}{0}
\setcounter{table}{0}
\setcounter{algocf}{0}
\renewcommand{\thesection}{S\arabic{section}}
\renewcommand{\thesubsection}{\thesection.\arabic{subsection}}
\renewcommand{\thetable}{S\arabic{table}}
\renewcommand{\thefigure}{S\arabic{figure}}
\renewcommand{\theequation}{S\arabic{equation}}
\renewcommand{\thealgocf}{S\arabic{algocf}}
\markboth{Zihan Xu}{Supplementary material}
\begin{center}
{\titlefont Supplementary material for\\Exact local queries for pseudo-marginal Markov chains\par}
\smallskip
Zihan Xu
\end{center}
\section{Additional proofs}
\label{supp:proofs}
\subsection{Summary of work bounds}
The notation, assumptions and numbered results cited below are those of
the main paper. Table~\ref{tab:work-bounds} collects the work guarantees.

\begin{table}[htbp]
\setlength{\tabcolsep}{4pt}
\renewcommand{\arraystretch}{1.08}
\newcommand{\workcell}[1]{{\raggedright #1\par}}
\tbl{Settings and work bounds for query execution}{%
\begin{tabular}{@{}p{.14\linewidth}p{.25\linewidth}%
                  p{.27\linewidth}p{.28\linewidth}@{}}
\workcell{Result} &
\workcell{Setting and executor} &
\workcell{Key conditions} &
\workcell{Work bound} \\[7pt]
\workcell{Theorem~\ref{thm:conditional-work}} &
\workcell{Finite graph; remaining queries from $\mathcal{C}_s$;
Algorithm~\ref{alg:query-execution}} &
\workcell{Canonical ranges; cost bounds measurable with respect to
$\mathcal{G}_s$} &
\workcell{Conditional bound for
$\mathrm{E}(W_s\mid\mathcal{G}_s)$ in
\eqref{eq:conditional-work-bound}} \\[7pt]
\workcell{Corollary~\ref{cor:bounded-degree-work}} &
\workcell{Bounded-degree graphs; at most $L$ updates per node;
Algorithm~\ref{alg:query-execution}} &
\workcell{Independent Gaussian proposals;
$(\Delta-1)q_D(w)<1$; common integrable $C$} &
\workcell{$\mathrm{E}\{W_{\mathrm{inc}}(Q)\}\le\kappa|Q|$,
uniformly in $n$} \\[7pt]
\workcell{Theorem~\ref{thm:path-block-work}} &
\workcell{Descending path blocks; $Q=\{(1,nL)\}$;
Algorithm~\ref{alg:path-block-execution}} &
\workcell{Independent Gaussian proposals; fixed finite $w,L$;
common integrable $C$} &
\workcell{$\mathrm{E}(N)\le(1-\rho)^{-1}$;
incremental work bounded by \eqref{eq:path-total-work},
uniformly in $n$} \\[7pt]
\workcell{Corollary~\ref{cor:pointwise-separation}} &
\workcell{The path and grid schedules in
\S\ref{sec:pointwise-comparison}; certificate execution and
$B_{\mathrm{point}}$} &
\workcell{Conditions of the corresponding family;
strictly positive estimates} &
\workcell{Expected incremental work:
$\Theta(nL)$ for $B_{\mathrm{point}}$;
$O(1)$ for certificate execution}
\end{tabular}}
\label{tab:work-bounds}
\begin{tabnote}
The table lists key conditions; full assumptions and operation costs
are given in the cited results.
Here $n$ is the number of graph vertices and $N$ counts resolved path blocks.
Bounds uniform in $n$ keep $L$, the input law and the applicable model
and cost parameters fixed.
Theorem~\ref{thm:conditional-work} concerns $W_s$, including the additional
preparation and output work in $b_s$; the other rows concern
$W_{\mathrm{inc}}$, with setup and output charged as in
\eqref{eq:query-work}.
\tablealt{Four results connect graph and query conditions to work bounds. On the specified paths and grids, expected incremental work stays bounded for certificate execution and grows linearly with graph size for the pointwise evaluator.}
\end{tabnote}
\end{table}

\subsection{Conditional work bound}
\label{supp:conditional-proof}
\begin{proof}[of Theorem~\ref{thm:conditional-work}]
We first establish the conditional law of the unrevealed thresholds.
Given $F$, the thresholds initially have a product uniform law.
At each revelation, the current transcript determines which index
is selected next, or whether revelation stops at the request boundary.
If an index is selected, conditioning on the corresponding threshold
leaves the other unrevealed coordinates independent and uniform.
The stopping decision uses only the current transcript and $F$,
so it imposes no further condition on those coordinates.
Induction over at most $T$ revelations gives the same product law
conditional on $\mathcal{G}_s$.
The cache is determined by $\mathcal{G}_s$, and every update with a
revealed threshold has returned and been cached at this boundary.
Hence the thresholds at the uncached updates form a subcollection
with this product law.

Fix $\mathcal{G}_s$ and list these updates as
$\tau_1<\cdots<\tau_m$ in the reference schedule.
All cached records are fixed under this conditioning.
The reference old record and candidate at $\tau_r$ can therefore
be reconstructed from $F$ and the earlier uncached thresholds.
The canonical endpoints at $\tau_r$ are consequently measurable
with respect to
$\mathcal{H}_{r-1}=\sigma(\mathcal{G}_s,U_{\tau_1},\ldots,
U_{\tau_{r-1}})$, where $\mathcal{H}_0=\mathcal{G}_s$.
Conditional on $\mathcal{H}_{r-1}$, the shift
$\underline{\alpha}_{\tau_r}$ is fixed and $U_{\tau_r}$ is uniform.
Translation modulo one preserves this law, so
\[
\operatorname{pr}(V_{\tau_r}\in A\mid\mathcal{H}_{r-1})
=\operatorname{Leb}(A),
\]
where $A$ is any Borel subset of $[0,1)$ and
$\operatorname{Leb}$ denotes Lebesgue measure.
Each earlier $V_{\tau_i}$, $i<r$, is
$\mathcal{H}_{r-1}$-measurable.
Successive conditioning therefore gives conditional independence
of all $V_{\tau_r}$, and hence of the events $B_t$, since each
$p_t$ is $\mathcal{G}_s$-measurable.

An unresolved neighbour is called only when the threshold lies in
the current undecided band.
Range refinement puts this band inside the canonical band.
Thus a call from update $t$ implies
$0\le V_t<\overline{\alpha}_t-\underline{\alpha}_t\le p_t$,
and hence $B_t$.
The same implication holds for every later call from this update,
which retains its original threshold.
The sets $\mathcal{E}_j$ are disjoint and
$\mathcal{G}_s$-measurable, so the node events $O_j$ are
conditionally independent with probabilities
\eqref{eq:node-open-probability}.

Consider first entries into nodes for uncached records.
An entry from a request gives a root in $\mathcal{Q}_s$;
otherwise, assign the calling node as parent.
A call for the preceding record at the calling node adds no edge.
Every parent is entered before its child, so these links form a forest.
Each edge requires an open parent by the preceding argument.
Whenever $I_j=1$, tracing parents therefore gives a path in
$\Gamma_s(q,j)$ for some $q\in\mathcal{Q}_s$, with every vertex
before $j$ open.
Conditional independence along each path and a union bound give
\eqref{eq:node-reachability-bound}.

The generation trigger condition and another union bound give
$\operatorname{pr}(J_j=1\mid\mathcal{G}_s)\le h_j^+$.
Taking conditional expectations in \eqref{eq:stage-work-accounting}
gives \eqref{eq:conditional-work-bound}, because all cost bounds
are $\mathcal{G}_s$-measurable.
The node bounds include work on own predecessors, and the generation
bounds include ranges at nodes whose records remain unresolved.
\end{proof}

\subsection{Path block bound}
\label{supp:path-proof}
\begin{proof}[of Theorem~\ref{thm:path-block-work}]
When the query returns, the resolved blocks form a prefix $1,\ldots,N$,
because a block can request only its right neighbour.
Write $p_i=p(F_i,F_{i+1})$ and $C_j=C(F_j)$.
For $0\le h<n$, let $M_h^{\mathrm{o}}$ and $M_h^{\mathrm{e}}$ be
the products of $p_i$ over the odd and even indices in
$1,\ldots,h$, respectively, with empty products equal to one.
Condition on all raw input blocks assigned to the node addresses.
Reaching block $h+1$ requires the first $h$ blocks to be canonically
open, so the conditional independence established in
\S\ref{supp:conditional-proof} gives
\[
\operatorname{pr}(N\ge h+1\mid F_1,\ldots,F_n)
\le\prod_{i=1}^{h}p_i.
\]
Averaging and applying Cauchy--Schwarz yields
\[
\begin{aligned}
\operatorname{pr}(N\ge h+1)
&\le\mathrm{E}(M_h^{\mathrm{o}}M_h^{\mathrm{e}})\\
&\le\bigl[\mathrm{E}\{(M_h^{\mathrm{o}})^2\}
          \mathrm{E}\{(M_h^{\mathrm{e}})^2\}\bigr]^{1/2}
=\rho^h.
\end{aligned}
\]
Each parity group uses disjoint pairs of input blocks, so its
second moment is a product of factors $\mathrm{E}(p_i^2)=\rho^2$.
There are $h$ factors across the two groups, which gives the final equality.
For $h\ge n$ the tail probability is zero, and summing the tail proves
\eqref{eq:path-block-tail}.

The candidate cost at block $h+1$ is bounded by
$C_{h+1}\mathbf{1}\{N\ge h+1\}$.
Since $C_{h+1}$ is determined by the input blocks, the conditional
bound above gives
\[
\mathrm{E}\{C_{h+1}\mathbf{1}\{N\ge h+1\}\}
\le\mathrm{E}(C_{h+1}M_h^{\mathrm{o}}M_h^{\mathrm{e}}).
\]
The expected candidate cost is at most $c$ at the root and at most
$\mathrm{E}(C_2p_1)=\delta$ at block 2, when present.
For $2\le h<n$, the variables $C_{h+1}^{1/2}M_h^{\mathrm{o}}$ and
$C_{h+1}^{1/2}M_h^{\mathrm{e}}$ have finite second moments because
their squares are bounded by $C_{h+1}$.
Cauchy--Schwarz therefore gives
\[
\begin{aligned}
\mathrm{E}(C_{h+1}M_h^{\mathrm{o}}M_h^{\mathrm{e}})
&\le\bigl[\mathrm{E}\{C_{h+1}(M_h^{\mathrm{o}})^2\}
          \mathrm{E}\{C_{h+1}(M_h^{\mathrm{e}})^2\}\bigr]^{1/2}\\
&=(c\varepsilon)^{1/2}\,\rho^{h-1}.
\end{aligned}
\]
In the group containing $p_h$, the final factor and endpoint cost contribute
$\mathrm{E}(C_{h+1}p_h^2)=\varepsilon$.
All other factors in that group use blocks disjoint from
$(F_h,F_{h+1})$.
The other parity group uses no $F_{h+1}$, so the cost contributes $c$.
The remaining $h-1$ factors each contribute $\rho^2$ before the square root.
Both $\delta$ and $\varepsilon$ are at most $c$, since $0\le p_i\le1$.
Summing over blocks gives
\[
\mathrm{E}\left(\sum_{j=1}^{N}C_j\right)
\le c+\delta+(c\varepsilon)^{1/2}
       \sum_{h=2}^{\infty}\rho^{h-1}
=c+\delta+\frac{\rho(c\varepsilon)^{1/2}}{1-\rho}.
\]

Let $G$ be the number of generated input blocks.
Only the immediate right neighbour of the resolved prefix can be
generated without being resolved, so $G\le\min\{n,N+1\}$.
Each resolved block requests its right neighbour at most once and
makes at most $L+1$ certificate comparisons.
The accounting in \S\ref{supp:counts} therefore bounds management
by $24L(N+1)$ logical units, including input generation and ranges
at the unresolved boundary block.
Adding its expectation to the candidate cost bound and using
\eqref{eq:path-block-tail} proves \eqref{eq:path-total-work}.
\end{proof}

\subsection{Same-query separation}
\label{supp:separation-proof}
\begin{proof}
The old parameter at an update is its initial value or an earlier
candidate at the same node.
Independent continuous proposals therefore differ from the old
parameter almost surely.
Together with strict positivity, this forces $B_{\mathrm{point}}$ to
resolve all neighbour records before each decision.
On the descending path schedule, resolving block $j<n$ requires the
last record of block $j+1$, so the query reaches every block.
On the grid's final scan, a vertex's west and north neighbours,
when present, precede it in the schedule.
Starting at the bottom-right vertex, these dependencies reach the
final-scan update of every vertex.
Resolving own predecessors then includes all $L$ updates at each node
in either family.
Caching evaluates each candidate once, giving
$N_{\mathrm{call}}^{\mathrm{point}}=nL$.
Each evaluation incurs at least one work unit.
The common envelope $C(F_j)$ bounds the remaining candidate work by
$\sum_{j=1}^n C(F_j)$, with expectation $nc$.
Bounded degree, indexed access and fixed $L$ give $O(nL)$ management
work, proving the stated order for $B_{\mathrm{point}}$.
The certificate bounds follow from
Theorem~\ref{thm:path-block-work} for paths and
Corollary~\ref{cor:bounded-degree-work} for grids, since $|Q|=1$.
Exact pointwise decisions and Theorem~\ref{thm:query-correctness}
give agreement with $\mathcal{R}_Q$.
\end{proof}

\subsection{Colour-class argument for Corollary~\ref{cor:bounded-degree-work}}
\label{supp:colour}
\begin{proof}[of Corollary~\ref{cor:bounded-degree-work}]
At the initial-record cache, $\mathcal{G}_0=\sigma(F)$.
For each node $j$, let $A_j$ be the value of $A$ formed from $F_j$.
The historical ranges allow the choice
$\omega_t=wA_j\sum_{k\in\mathcal{N}_j}A_k$ at each update of node $j$.
There are at most $L$ such updates, so
\[
\pi_j\le\bar\pi_j
:=1-\exp\left(-wLA_j\sum_{k\in\mathcal{N}_j}A_k\right).
\]
Write $q=q_D(w)$.
Padding the neighbour sum to $\Delta$ terms with independent copies of $A$ gives
$\mathrm{E}(\bar\pi_j^D)\le q^D$.

Let $\mathcal{K}_1,\ldots,\mathcal{K}_D$ be the colour classes.
Distinct vertices in one class have disjoint closed neighbourhoods
$\{j\}\cup\mathcal{N}_j$,
so their $\bar\pi_j$ depend on disjoint input blocks and are independent.
For any vertex set $\mathcal{A}\subseteq V$, set
$\mathcal{A}_d=\mathcal{A}\cap\mathcal{K}_d$.
H\"older's inequality across classes and independence within each class give
\[
\mathrm{E}\left(\prod_{v\in\mathcal{A}}\bar\pi_v\right)
\le\prod_{d=1}^{D}
\left\{\mathrm{E}\left(\prod_{v\in\mathcal{A}_d}\bar\pi_v^D\right)\right\}^{1/D}
=\prod_{v\in\mathcal{A}}
\{\mathrm{E}(\bar\pi_v^D)\}^{1/D}
\le q^{|\mathcal{A}|}.
\]
Empty products equal one.

For a self-avoiding path $\gamma=(v_0,\ldots,v_\ell)$, let
$\mathcal{P}_\gamma=\{v_i:0\le i<\ell\}$ be its vertices
before the endpoint.
The conditional reachability bound in
\eqref{eq:node-reachability-bound} assigns this path weight
$\prod_{v\in\mathcal{P}_\gamma}\pi_v$, whose expectation is at most
$q^\ell$ by the preceding inequality.
At most $\Delta(\Delta-1)^{\ell-1}$ paths of length $\ell\ge1$
start at a given vertex.
For $a\in\{0,\Delta\}$, define
\[
\Xi_a=1+\sum_{\ell=1}^{\infty}
\Delta(\Delta-1)^{\ell-1}q^{(\ell-a)_+},
\]
where $x_+=\max\{x,0\}$.
Both series converge because their tails have ratio
$(\Delta-1)q<1$.
Let $I_j$ indicate that at least one new record at $j$ is resolved.
Summing the reachability bound over endpoints and the at most $|Q|$
query nodes gives
\[
\mathrm{E}\left(\sum_j I_j\right)\le |Q|\Xi_0.
\]

For a path ending at $j$, remove all neighbours of $j$ from
$\mathcal{P}_\gamma$ and write
$\mathcal{P}'_\gamma=\mathcal{P}_\gamma\setminus\mathcal{N}_j$.
This removes at most $\Delta$ vertices.
Each remaining factor $\bar\pi_v$ depends only on blocks other than
$F_j$, so their joint product is independent of $C(F_j)$.
Since $0\le\bar\pi_v\le1$, we obtain
\[
\begin{aligned}
\mathrm{E}\left\{C(F_j)
\prod_{v\in\mathcal{P}_\gamma}\bar\pi_v\right\}
&\le\mathrm{E}\left\{C(F_j)
\prod_{v\in\mathcal{P}'_\gamma}\bar\pi_v\right\}\\
&=c\,\mathrm{E}\left(\prod_{v\in\mathcal{P}'_\gamma}\bar\pi_v\right)
\le c q^{|\mathcal{P}'_\gamma|}
\le c q^{(\ell-\Delta)_+}.
\end{aligned}
\]
Here $|\mathcal{P}'_\gamma|\ge(\ell-\Delta)_+$ and $0<q<1$.
The cost $C(F_j)$ is $\mathcal{G}_0$-measurable, so we may multiply
the conditional reachability bound by it before taking expectations.
Summing the resulting path bounds gives
\[
\mathrm{E}\left(\sum_j C(F_j)I_j\right)\le c|Q|\Xi_\Delta.
\]

It remains to account for management work outside the candidate cost bounds.
At most $L$ updates are resolved at each entered node.
Each update makes at most $\Delta+1$ comparisons and requests each
unresolved neighbour at most once.
With per-node event indices, each historical lookup or range calculation
examines at most $L$ entries per neighbour.
Charge each range calculation to the update that requests it, including
ranges at nodes whose records remain unresolved.
The raw inputs and event indices are already prepared under the
corollary's cost convention.
Candidate generation and evaluation are covered by $C(F_j)$.
The remaining management work is therefore at most
$K(|Q|+\sum_j I_j)$, where $K<\infty$ depends only on $\Delta,L$
and the fixed operation costs.
Combining the bounds gives
\[
\mathrm{E}\{W_{\mathrm{inc}}(Q)\}
\le\{c\Xi_\Delta+K(1+\Xi_0)\}|Q|,
\]
which proves the corollary with
$\kappa=c\Xi_\Delta+K(1+\Xi_0)$.
\end{proof}

\section{Block procedure and operation accounting}
\label{supp:block}
\subsection{Block execution}
Algorithm~\ref{alg:path-block-execution} specializes the executor to the
descending path schedule of \S\ref{sec:path-blocks} in the main paper.
At the reference times in block $j$, its left neighbour is still at its
initial state. Its right neighbour is at its block-end state.
Recursive execution uses these historical states even when the caller
has already advanced to a different state.
The initial cache in the first line belongs to preparation. Input blocks
and their ranges are generated when requested, and repeated addresses
return the same random input.

\begin{algorithm}[tbp]
\caption{Block query execution on a path.}
\label{alg:path-block-execution}
\SetKwProg{Fn}{Procedure}{:}{}
\SetKwFunction{GetBlock}{GetBlock}
\SetKwFunction{ResolveBlock}{ResolveBlock}
\LinesNumbered
\KwIn{The path specification and addressable inputs in
\S\ref{sec:path-blocks}.}
\KwOut{The complete record $R_1^{(nL)}$.}
Initialize $\mathcal{C}[j,0]=R_j^{(0)}$, $j=1,\ldots,n$, and
empty caches for input blocks and their intervals\;
\Fn{\GetBlock{$j$}}{
  \If{the input block at $j$ is not cached}{
    Generate and cache its $L$ parameter and raw auxiliary inputs\;
    Cache the interval spanned by $0,Y_{j1},\ldots,Y_{jL}$\;
  }
}
\Fn{\ResolveBlock{$j$}}{
  $\tau\gets(n-j+1)L$\;
  \lIf{$(j,\tau)$ is cached}{\KwRet $\mathcal{C}[j,\tau]$}
  \GetBlock{$j$}; $R\gets\mathcal{C}[j,0]$\;
  \For{$\ell\gets1$ \KwTo $L$}{
    $t\gets(n-j)L+\ell$\;
    Form and evaluate
    $R^+=(Y_{j\ell},v,\widehat{\ell}_j(Y_{j\ell},v))$
    once from the cached inputs at $(j,\ell)$\;
    \If{$\widehat{\ell}_j(Y_{j\ell},v)>0$}{
      Retrieve and retain $U_t$ from its address\;
      Use the singleton $\{0\}$ for the left neighbour when $j>1$\;
      \If{$j<n$}{
        \eIf{$(j+1,(n-j)L)$ is cached}{
          Use its parameter as the right singleton\;
        }{
          \GetBlock{$j+1$}; use its cached interval on the right\;
        }
      }
      Compute \eqref{eq:log-ratio-bounds} and apply
      \eqref{eq:decision-certificates} to $R,R^+,U_t$\;
      \If{the comparison is undecided}{
        \ResolveBlock{$j+1$}; repeat the comparison using its returned
        parameter on the right, retaining $R,R^+,U_t$\;
      }
      \lIf{the comparison accepts}{$R\gets R^+$}
    }
    $\mathcal{C}[j,t]\gets R$\;
  }
  \KwRet $R$\;
}
\KwRet \ResolveBlock{$1$}\;
\end{algorithm}

\subsection{Logical work units}
\label{supp:counts}
Let $G$ be the number of generated input blocks and $N$ the number
of resolved blocks.
At query completion, the resolved blocks form a prefix, and only its
immediate right neighbour can be generated without being resolved.
Thus $N\le G\le\min\{n,N+1\}$.
Each resolved block requests its right neighbour at most once.
After that request returns, the neighbour's block-end record is
cached for all remaining updates of the calling block.
A block therefore makes at most $L$ initial comparisons and one
comparison after recursion returns.
A zero candidate estimate is rejected before its threshold is retrieved.
Each generated block supplies $L$ parameter inputs and $L$ raw
auxiliary inputs, and its range is formed once from $L+1$ values.
Repeated accesses use the cached inputs and range.
Table~\ref{supp:count-table} counts candidate dispatch separately
from the construction and evaluation work covered by $C(F_j)$.
A logical unit includes the fixed scalar arithmetic and record access
needed for the stated operation.
For range construction, one logical unit covers both endpoint comparisons
for one value.
\begin{table}[tbp]
\tbl{Operation counts for the path implementation}{%
\begin{tabular}{ll}
Operation & Upper bound\\[3pt]
Scheduling & $LN+2N$\\
Certificate comparisons & $LN+N$\\
Stack push, pop and recursive request & $3N$\\
Candidate dispatch, commit and rejection handling & $3LN$\\
Block lookup & $N+LN$\\
Threshold read & $LN$\\
Parameter and raw auxiliary input generation & $2LG$\\
Range element visits & $(L+1)G$\\
\end{tabular}}
\label{supp:count-table}
\begin{tabnote}
These are logical operation units for the specified implementation.
Fixed-size records, indexed access and amortized cache operations have
unit cost. Backend evaluation, initial records and data preparation
retain their separate charges.
\tablealt{Scheduling and certificate comparisons scale with resolved blocks and their length. Input generation and range visits depend on how many blocks have their inputs generated.}
\end{tabnote}
\end{table}
The first six rows sum to at most $7LN+7N$ logical units.
The final two rows add at most $(3L+1)G$ units.
Since $L\ge1$ and $G\le N+1$, the total satisfies
\[
\begin{aligned}
7LN+7N+(3L+1)G
&\le18LN+4L\\
&\le18L(N+1)\le24L(N+1).
\end{aligned}
\]
With other fixed operation prices, multiply this upper bound by
their largest price.
These are costs of the listed execution operations, with preparation
and statistical outputs charged separately as in \eqref{eq:query-work}.

For the experiment with estimates that can be zero, consider $w=0.01,0.2,1.5$ in that order.
The mean certificate candidate counts are 1.25 at all three weights for $L=1$, and 2.5, 2.5 and 3.5 for $L=2$.
For $L=4$, the corresponding means are 5, 9 and 12.
Each mean uses the same four fixed input streams and is unchanged across $n\in\{4,16,64,256\}$.

For the positive-estimate experiment, we set $R_j^{(0)}=(0,0,e^{-1/2})$ at every node and store the estimates on the log scale.
For $u\sim N(0,1)$, the identity $\mathrm{E}\{\exp(u-1/2)\}=1$ gives unbiasedness for $\ell_j(\theta)=\exp(-\theta^2/2)$.
For each input stream, parameter candidates, auxiliary inputs and acceptance thresholds are identical at common node and update addresses across the four path lengths.
Each executor prepares its own initial records and starts with a fresh cache.

For every input stream, the number of resolved blocks is unchanged across the four path lengths.
At each path length, the sample standard deviation of $N$ is 0.807.
In all 400 certificate runs, $G=N+1$.
Management uses 54--281 logical units and satisfies the bound $24L(N+1)$ in every run.

Table~\ref{supp:positive-path-times} reports mean times for certificate execution in the instrumented implementation.
Each row uses the same 100 input streams.
Incremental time includes input generation, range maintenance and candidate evaluation.
Total time also includes preparation, query output and time outside the separately timed stages.
The accompanying files \texttt{data/positive\_path\_timings.csv} and \texttt{data/positive\_path\_summary.csv} report the individual runs and their summaries.

\begin{table}[tbp]
\tbl{Mean certificate execution times (ms) for the positive-estimate path experiment}{%
\begin{tabular}{rrrr}
$n$ & Preparation & Incremental & Total\\[3pt]
32   & 0.023 & 0.420 & 0.451\\
128  & 0.079 & 0.430 & 0.517\\
512  & 0.249 & 0.389 & 0.645\\
2048 & 1.371 & 0.490 & 1.870
\end{tabular}}
\label{supp:positive-path-times}
\begin{tabnote}
\tablealt{As path size increases from 32 to 2048 nodes, mean preparation time rises from 0.023 to 1.371 milliseconds. Mean incremental time stays between 0.389 and 0.490 milliseconds.}
\end{tabnote}
\end{table}

\section{Controlled experiments}
\label{supp:experiments}
\subsection{Models and fixed configurations}
The six configurations in Figure~\ref{fig:same-query-cost} use four saved simulated datasets.
Figure~\ref{fig:query-scope-cost} uses P1 and P2.
All have 128 scalar parameters on a path with node precision one and
edge weight $0.8$. The parameter prior is the Gaussian graph prior in
\eqref{eq:graph-posterior}. P1 and P2 use datasets 0 and 1. S1 and B1
use dataset 8; S2 and B2 use dataset 9. Within each gene-activity dataset, the S and B configurations use different particle estimators and parameter proposals.
The data and statistical model are the same within each pair.

For P1 and P2, the latent process and observations are
\[
\begin{aligned}
X_{j1}&\sim N\{0,0.4^2/(1-0.7^2)\},\qquad
X_{jt}\mid X_{j,t-1}\sim N(0.7X_{j,t-1},0.4^2),\\
Y_{jt}\mid X_{jt},\theta_j&\sim\mathrm{Poisson}\{8e^{\theta_j+X_{jt}}\},
\end{aligned}
\]
where $X_{jt}$ is the latent state and $Y_{jt}$ is the observed count at
node $j$ and observation time $t$. There are 16 observations per node.
A bootstrap particle filter uses 64 particles with multinomial
resampling at each subsequent observation. The parameter proposal is
$N(\theta_j,0.35^2)$ and the schedule contains $128\times192$ updates.

For S1, S2, B1 and B2, a promoter $G_j(t)\in\{0,1\}$ switches on at
rate $0.7e^{\theta_j}$ and off at rate $0.9$. Molecules are produced at
rate $3G_j(t)$ and each degrades at rate $0.8$.
The promoter and molecule count start at zero. Counts are observed at
eight times separated by $0.5$ time units, with independent capture
probability $0.95$ for each molecule present.
Conditional on a promoter holding segment, survivors are binomial and
new surviving molecules are Poisson. The implementation simulates
these transitions without truncating the molecule count.

B1 and B2 use a bootstrap filter with 16 particles and a
$N(\theta_j,0.35^2)$ parameter proposal. S1 and S2 use 16 particles
with observation-conditioned count endpoints and an independent
$N(0,1)$ parameter proposal. Given the promoter path during an interval,
the observed survivors are the sum of independent binomial and Poisson
variables. Their finite convolution supplies the observation weight;
the unobserved survivors are then drawn conditionally on the observation.
When the promoter starts off and the observed count exceeds the initial
molecule count, the first activation is conditioned to occur within the
interval. Its probability is included in the importance weight.
All four configurations use $128\times48$ updates. The normalized
parameter proposal density is included in the acceptance ratio.
All particle estimates use products of average incremental weights
\citep{AndrieuDoucetHolenstein2010}.

Each comparison shares initial complete records, node-selection inputs,
parameter inputs, particle seeds and acceptance thresholds. Initial
estimates are obtained by the same initialization procedure and are
charged in each execution. The sparse task returns the two node histories
at every reference time. The full task returns all 128 histories.
The five sparse repeats and the single full-query repeat are the fixed
timing design, with executor order cyclically rotated across repeats.
They are not additional independent datasets.

The cache comparison uses 128 saved trajectories from eight Poisson datasets and eight gene-activity datasets, with eight runs per dataset.
We include all eight designated runs for each dataset.
Each run uses the 128-node path prior specified above and requests the complete parameter histories of nodes 1 and 2.
Six Poisson datasets use exposure 8 and two use exposure 1.
Six gene-activity datasets use capture probability $0.95$ and two use $0.65$.
The remaining model parameters are as specified above.
The Poisson runs use 64 bootstrap particles, $N(\theta_j,0.35^2)$ parameter proposals and $128\times192$ updates.
The gene-activity runs use 16 particles with observation-conditioned count endpoints, independent $N(0,1)$ parameter proposals and $128\times48$ updates.

For each decision made with at least one required neighbour record unresolved, we retain the old record, candidate record and acceptance threshold.
We propagate historical ranges from the initial states using the same proposal inputs and recompute the acceptance bounds.
We count a decision as aided by cached ranges if these replacement bounds leave it undecided.
A decision enters this count whether it occurs at the first test or after earlier neighbour requests.

\subsection{Measured costs and individual runs}
\label{supp:timings}
The experiments used Windows 11, an Intel64 Family 6 Model 183 processor
with 24 logical processors, Python 3.12.14 and NumPy 2.3.5.
The runs used the same host in sequential deployments. Four likelihood
workers were available for common initialization and for Picard
candidate evaluation. Certificate and pointwise production updates
were evaluated in the controlling process.
The Picard window is 16. Numerical-library thread limits were one.
These comparisons measure task latency; worker job times can overlap.

Elapsed time is the saved algorithm time plus the input loading charge.
It includes initialization, worker startup and shutdown, random inputs,
indices, query execution and the prescribed summary calculations.
Simulated-data generation, result auditing and audit serialization are
outside this production timer. The accompanying
\texttt{data/execution\_timings.csv} contains all 180 recorded deployments,
including executor, query scope, repeat, candidate count and timing
components. Tables~\ref{supp:time-summary}--\ref{supp:scope-summary}
summarize the comparisons used in the main paper. The per-run paired
records are supplied as \texttt{data/paired\_timings.csv}.

\begin{table}[tbp]
\tbl{Sparse queries: measured elapsed time in seconds}{%
\begin{tabular}{lrrrr}
Case & Pointwise & Certificate & Mean ratio & Paired ratio range\\[4pt]
P1 & 12.995 (6.837) & 2.613 (1.227) & 0.201 & 0.102--0.319\\
P2 & 18.888 (1.409) & 3.727 (0.893) & 0.197 & 0.151--0.245\\
S1 & 7.996 (1.370) & 3.030 (0.881) & 0.379 & 0.216--0.504\\
S2 & 6.699 (1.104) & 3.155 (0.302) & 0.471 & 0.389--0.643\\
B1 & 5.022 (1.859) & 1.529 (0.397) & 0.304 & 0.277--0.385\\
B2 & 2.725 (1.198) & 1.300 (0.287) & 0.477 & 0.254--0.874\\
\end{tabular}}
\label{supp:time-summary}
\begin{tabnote}
Parentheses give standard deviations over five runs.
The mean ratio is mean certificate time divided by mean pointwise time.
The paired ratio range gives the minimum and maximum certificate-to-pointwise time ratios across the five runs.
\tablealt{Certificate execution is faster than pointwise execution in every paired run. Across the six cases, its mean elapsed time ranges from 20\% to 48\% of the pointwise mean.}
\end{tabnote}
\end{table}
\begin{table}[tbp]
\tbl{Query scope and comparison with Picard execution}{%
\begin{tabular}{lrrrrrr}
&\multicolumn{3}{c}{Sparse queries: five-run means}&\multicolumn{3}{c}{All queries: one run}\\Case & Certificate & Picard & Ratio & Certificate & Picard & Ratio\\[4pt]
P1 & 2.613 & 8.370 & 0.312 & 28.035 & 12.275 & 2.284\\
P2 & 3.727 & 10.945 & 0.341 & 25.697 & 10.373 & 2.477\\
S1 & 3.030 & 7.125 & 0.425 & 32.094 & 7.402 & 4.336\\
S2 & 3.155 & 7.287 & 0.433 & 34.309 & 7.587 & 4.522\\
B1 & 1.529 & 4.157 & 0.368 & 22.988 & 4.286 & 5.363\\
B2 & 1.300 & 4.187 & 0.310 & 19.014 & 4.600 & 4.134\\
\end{tabular}}
\label{supp:scope-summary}
\begin{tabnote}
All times are seconds. The sparse certificate runs are the same runs as in Table~\ref{supp:time-summary}. Each ratio uses certificate time divided by Picard time.
\tablealt{In all six cases, certificate execution has a shorter mean time than Picard for sparse queries. Picard is faster in each single run covering all queries.}
\end{tabnote}
\end{table}

For the two-node task in these six configurations, mean resolved fractions for certificate execution range from 2.69\% to 11.95\%.
For the full task, all scheduled updates are resolved.
Within each configuration, the two scopes use the same initial records and random inputs in the first paired run.
Picard's candidate evaluation counts and parallel rounds agree between the two scopes in all six such comparisons.

The additional scope experiment uses P1 and P2 with the model, particle filter and proposal settings given above.
We request the histories of nodes 1 through $m$, where $m\in\{2,8,32,64,128\}$.
For each dataset, we use five random-input streams at every scope.
Within each stream, all three executors share the initial complete records and random inputs at every scope.
Each execution pays for initialization and starts with a fresh cache.
The Picard window is fixed at 16 updates, with four likelihood workers.

The requested histories agree across executors for each dataset, input stream and query scope.
At 64 nodes, the individual ratios of certificate time to Picard time range from 0.948 to 1.037 in P1 and from 0.987 to 1.050 in P2.
The accompanying \texttt{data/query\_scope\_timings.csv} contains all 150 deployments.
The file \texttt{data/query\_scope\_summary.csv} gives the mean costs and their standard deviations.

For the window check, we use the first input stream and request either two node histories or all 128.
For each dataset and query scope, we run Picard once at each window size of 4, 16 and 64 updates.
With four workers, increasing the window from 16 to 64 raises candidate calls by 4.8\% in P1 and 5.3\% in P2.
The number of parallel rounds falls from 2,955 to 842 in P1 and from 2,951 to 849 in P2.
Picard time decreases in all four comparisons.
The accompanying \texttt{data/query\_window\_timings.csv} contains all 12 executions.

\section{Application specifications and diagnostics}
\subsection{Tomography model and numerical target}
\label{supp:pet}
The input is subject 01, baseline session, in OpenNeuro ds004869, version 1.4.1 \citep{YanEtAl2025MC1,YanEtAl2026MC1Data}.
We use the first 27 frames, covering 90\,min, together with the released plasma and whole-blood curves and observation weights.
The dataset is released under the Creative Commons CC0 public domain dedication.
The original human study received approval from the National Institutes of Health
Combined Neurosciences Institutional Review Board under protocols
19-M-0079 and 20-M-0082.

We map the released anatomical labels to the PET image grid using the supplied spatial transform and nearest-neighbour interpolation.
Label 1035 (left insula), restricted to the cortical brain mask, defines the 1311 query voxels.
We connect pairs of face-sharing voxels in the left cortical brain mask.
The analysis domain consists of the connected components containing query voxels and has 41\,500 voxels.

Activity concentrations are expressed in $\mathrm{kBq}\,\mathrm{mL}^{-1}$ and time in minutes.
The convolution grid has 6000 equally spaced points.
We subtract a delay of $-0.12912561648752363$ minutes from the frame midpoints to obtain the evaluation times.
The blood fraction is fixed at $v_b=0.05$.

For a one-tissue model the tissue curve is
$K_1(C_p*e^{-k_2\cdot})(t)$. For a two-tissue model it is
$C_p*(c_f e^{-\lambda_f\cdot}+c_s e^{-\lambda_s\cdot})(t)$, where
$C_p$ is the plasma curve, $*$ denotes convolution,
$\lambda_f=(k_2+k_3+k_4+g)/2$,
$\lambda_s=2k_2k_4/(k_2+k_3+k_4+g)$,
$g=\{(k_2-k_4)^2+k_3^2+2k_3(k_2+k_4)\}^{1/2}$,
$c_f=K_1(\lambda_f-k_3-k_4)/g$ and
$c_s=K_1(k_3+k_4-\lambda_s)/g$.
The measured curve combines $(1-v_b)$ times the tissue curve with
$v_b$ times the whole-blood curve. The code uses rectangle-sum
convolution on the fixed grid and linear interpolation to the frame
midpoints.

Given the kinetic parameters and model label, observations have
independent Gaussian errors with precision $w_t\tau_j$, where $w_t$
is the supplied weight and $\tau_j\sim\mathrm{Gamma}(2,1)$ uses a rate
parameterization. The kinetic parameters have independent uniform
priors on their logarithms. The original-scale bounds are
$K_1,k_2,k_3\in[0.0001,0.5]$ and $k_4\in[0.01,0.2]$; the one-tissue
model uses only $K_1,k_2$. Here $K_1$ is in
$\mathrm{mL}\,\mathrm{cm}^{-3}\,\mathrm{min}^{-1}$ and the other rates
are in $\mathrm{min}^{-1}$.
Distribution volume is $K_1/k_2$ for the one-tissue model and
$(K_1/k_2)(1+k_3/k_4)$ for the two-tissue model.

Integrating the noise precision analytically gives a likelihood proportional to $(1+\mathrm{RSS}/2)^{-(2+27/2)}$, where $\mathrm{RSS}$ is the weighted residual sum of squares.
The two-tissue model uses 96 Gauss--Legendre points in each of $\log k_2$, $\log k_3$ and $\log k_4$, and 32 points in each positive-width $\log K_1$ segment.
The one-tissue model uses 64 points in each positive-width $\log K_1$ segment.
For its outer $\log k_2$ integral, adaptive Gauss--Kronrod integration determines a partition of the prior interval.
A 32-point Gauss--Legendre rule on each final subinterval supplies the positive quadrature weights used for the evidence and volume distribution.
The partition is computed for fixed groups of voxels, with group membership specified before queries.

For both models, the $K_1$ segment boundaries are the prior endpoints and the clipped values of a conditional least-squares centre plus scale times $-256,-64,-16,-4,-1,0,1,4,16,64,256$.
Taking logarithms gives the $\log K_1$ segments, and segments of zero width contribute zero.
The centre, scale and numerical rules are specified in \texttt{specifications/pet/integration.py} and \texttt{specifications/pet/quadrature.py}.
All partitions and quadrature values are deterministic functions of the fixed data, groups and numerical settings.
Every positive atom is assigned to a fixed logarithmic volume bin, of log width at most $0.002$, and represented by the geometric midpoint of that bin.

The graph has $n=41500$ voxels and $85619$ face-adjacent edges.
Write $b=(b_1,\ldots,b_n)\in\{0,1\}^n$ for the model labels, with $0$ denoting one tissue compartment and $1$ denoting two.
The numerical label target is
\[
\widetilde\pi_B(b)\propto \left\{\prod_{j=1}^n\widetilde Z_j(b_j)\right\} \exp\left\{\beta\sum_{\{j,k\}\in E}\mathbf{1}\{b_j=b_k\}\right\},
\]
where $\widetilde Z_j(k)>0$ is the deterministic quadrature evidence for voxel $j$ and label $k$, $E$ is the edge set, $\beta=0.25$, and $\mathbf{1}\{\cdot\}$ denotes an indicator.
The equal label base probabilities cancel from this expression.

Each reference update selects a voxel uniformly and proposes label $0$ or $1$ with equal probability, independently of the current labels.
At update $t$ at voxel $j$, write $x$ and $y$ for the old and proposed labels and $U_t\in(0,1]$ for the independent uniform acceptance input.
If $y=x$, the log acceptance ratio is zero and both executions return $x$ without evaluating model evidence or resolving a neighbour label.
Otherwise, symmetry of the proposal gives the log acceptance ratio
\[
\Lambda_t=\log\widetilde Z_j(y)-\log\widetilde Z_j(x) +\beta\sum_{k:\{j,k\}\in E}\bigl[\mathbf{1}\{y=b_k\}-\mathbf{1}\{x=b_k\}\bigr],
\]
where $b_k$ is the neighbour label immediately before update $t$.

For $y\ne x$, let $K_t$ be the set of neighbours with the required historical label already cached at the current comparison.
Write $r_t$ for the number of remaining neighbours.
Define
\[
a_t=\log\widetilde Z_j(y)-\log\widetilde Z_j(x) +\beta\sum_{k\in K_t}\bigl[\mathbf{1}\{y=b_k\}-\mathbf{1}\{x=b_k\}\bigr]. \]
Each unresolved neighbour contributes either $-\beta$ or $\beta$, so $\Lambda_t\in[a_t-\beta r_t,a_t+\beta r_t]$.
The update is accepted if $\log U_t<\min\{0,a_t-\beta r_t\}$ and rejected if $\log U_t\ge\min\{0,a_t+\beta r_t\}$.
An undecided comparison resolves one unresolved neighbour at the required earlier time and repeats the test with the same proposal and uniform variable.

Label records are cached by voxel and update time; each call resolves the voxel's own predecessor before applying the test.
Evidence values are deterministic and are cached separately by voxel and model label.
Fix a finite update schedule, a finite ordered query list, the initial labels and all random inputs.
Work in exact arithmetic and assume that every required quadrature table can be evaluated in finite time.
Every recursive call requests an earlier historical label, and each undecided comparison resolves a previously unresolved neighbour.
Once all neighbours are resolved, the interval is a singleton and the test decides the update, including equality at the threshold.
Given correct cached labels, the interval contains the reference log ratio, so each certificate gives the reference decision.
Induction over update times therefore proves termination and agreement with the full reference labels.
The symmetric Metropolis rule preserves $\widetilde\pi_B$, and uniform selection of the updated voxel preserves it as well.

For voxel $j$ and label $k$, define the conditional volume distribution
\[
H_{jk}(\mathrm{d}v)=\sum_r p_{jkr}\,\delta_{v_{jkr}}(\mathrm{d}v),
\]
where $r$ indexes the fixed bins, $p_{jkr}$ is the normalized quadrature mass in bin $r$, $v_{jkr}$ is its geometric midpoint, and $\delta_v$ is unit mass at $v$.
The joint numerical target for the labels and regional volumes is
\[
\widetilde\Pi(b,\mathrm{d}v_R)=\widetilde\pi_B(b)\prod_{j\in R}H_{j,b_j}(\mathrm{d}v_j),
\]
where $R$ is the 1311-voxel left-insula region and $v_R=(v_j)_{j\in R}$.
At each query time, independent output uniforms generate $v_j\sim H_{j,b_j}$ for $j\in R$, and the regional draw is $|R|^{-1}\sum_{j\in R}v_j$.
Identical requested labels and output uniforms give identical regional draws.
Averaging the conditional regional means estimates the posterior mean; empirical quantiles of the regional draws estimate the interval endpoints.
Monte Carlo error is assessed relative to this fixed numerical target.

We run four chains for 512 nominal sweeps each.
A nominal sweep consists of 41,500 random-scan updates.
The four initial label vectors are all zero, all one, alternating zero and one in the fixed voxel order, and the reverse alternation.
We request the regional output every 10,375 updates, giving 2048 queries per chain.
Discarding the first 256 queries (64 sweeps) leaves 1792 label vectors per chain and 7168 in total.

We use the saved conditional moments of the quadrature distributions to assess the effect of replacing support points by bin midpoints.
The evidence values and the four saved label chains remain fixed.
Using the unbinned conditional moments decreases the estimated regional mean from $2.0537176909$ to $2.0537175120\,\mathrm{mL}\,\mathrm{cm}^{-3}$.
The corresponding estimate of total posterior variance decreases by $0.0033\%$.

We also use the lower and upper endpoints of each sampled bin to bound changes in the empirical regional quantiles.
For the saved post-warmup draws, the resulting bounds for the empirical 2.5\% quantile are $[2.04042,2.04451]\,\mathrm{mL}\,\mathrm{cm}^{-3}$.
The corresponding bounds for the empirical 97.5\% quantile are $[2.06261,2.06675]\,\mathrm{mL}\,\mathrm{cm}^{-3}$.
These bounds concern bin placement for the saved draws and exclude Monte Carlo uncertainty and quadrature error.

\begin{figure}[tbp]
\centering
\setlength{\figwidth}{\linewidth}
\includegraphics[width=\figwidth]{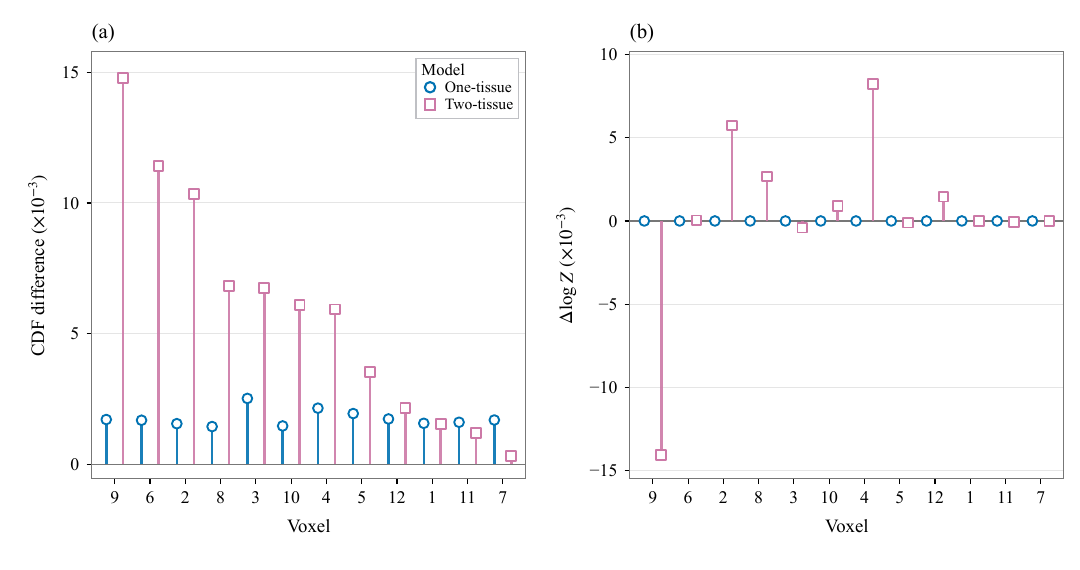}
\caption{Sensitivity to finite quadrature rules at 12 prespecified left-insula voxels.
(a) Maximum absolute difference between the cumulative distribution functions (CDFs) of the binned conditional $V_T$ distributions under the two rules.
(b) Change in log evidence, $\Delta\log Z=\log Z_{\mathrm{high}}-\log Z_{\mathrm{low}}$, where $Z_{\mathrm{low}}$ and $Z_{\mathrm{high}}$ denote the corresponding quadrature evidence values.
For the one-tissue model, the inner Gauss--Legendre order increases from 32 to 64.
For the two-tissue model, each outer order increases from 48 to 96, with inner order 32.
Voxel indices identify the 12 voxels in their saved order.
Both panels sort these indices by decreasing two-tissue CDF difference.
The legend applies to both panels.
Each vertical segment joins its value to zero.}
\label{supp:pet-quadrature}
\figurealt{For the two-tissue model, voxel 9 shows the largest difference between cumulative distributions in panel (a) and the largest drop in log evidence in panel (b). One-tissue log evidence changes little across all 12 voxels.}
\end{figure}

Figure~\ref{supp:pet-quadrature} compares conditional cumulative distribution functions (CDFs) and log evidence under two finite quadrature rules at 12 prespecified regional voxels.
For the two-tissue model, voxel~9 has the largest CDF difference and the largest absolute change in log evidence.
At voxel~6, the two-tissue CDF difference is $0.01141$.
The corresponding log evidence increases by $3.73\times10^{-5}$.
The one-tissue log evidence changes by at most $1.01\times10^{-12}$ across these voxels.

\subsection{Tomography cost accounting}
\label{supp:pet-cost}
Both compared executors start from the same 2942 cached voxel--model
evidence values and the regional conditional distributions.
We time each process from launch to exit, including cache loading, random input generation and output writing.
The full executor uses the corresponding optimized evidence calculations,
cached values and a current-state representation. Its budget ends at
725 updates, before the first requested output at event 10375.
It completes 353 new two-tissue evidence calculations, compared with
177 in certificate execution. The measured task times are 45.03 and
23.29 minutes, respectively; only certificate execution completes the
2048 prescribed queries.

The historical components traceable to the initial cache total about
7.01 hours: 6.35 hours for regional integration, 0.65 hours for
integration outside the region, and about 56 seconds for basis
preparation. These are deduplicated historical components, not a measured
cold start of the final implementation. Group construction charges
already included in these components are counted once.
The recorded task comparison therefore measures execution from an
existing cache. The full task's completion time is unobserved.
The accompanying preparation-component ledger identifies the source
events used for this accounting.

\subsection{Landmine model and sampling schedules}
\label{supp:landmine}
We use \texttt{LandmineData.mat} from the public mirror at
\url{https://github.com/daizhongxiang/Differentially-Private-Federated-Bayesian-Optimization/blob/main/LandmineData.mat}.
Sites 1--15 are the original highly foliated group \citep{XueEtAl2007}.
We choose training indices without using labels.
Within each site, we randomly order groups of identical feature rows and retain whole groups until at least 200 rows have been selected.
The saved training set has exactly 200 rows per site.
Features are centred and scaled using the pooled training mean and population standard deviation.
The fixed indices and scaling inputs accompany the specifications.

Each binary label, encoded as $y\in\{-1,1\}$, has conditional likelihood $\Phi(yf)$, where $\Phi$ is the standard normal distribution function and $f$ is the corresponding latent value.
At site $j$, the latent Gaussian covariance has entries
\[
K_{j,rs}=e^{2\gamma_j}\exp\{-\|x_{jr}-x_{js}\|^2/(2e^{2z_j})\}+4+10^{-6}\mathbf{1}_{\{r=s\}},
\] where $x_{jr}$ is the standardized feature vector of training row $r$, $z_j$ is the log length scale and $\gamma_j$ is the log signal standard deviation.
The constant term is the variance of a common intercept within the site, and the final term is added only to diagonal entries.
The log length scales are conditionally independent given $\mu$, with $z_j\mid\mu\sim N(\mu,\sigma^2)$ and $\mu\sim N(\log3,\sigma^2)$, where $\sigma^2=0.75^2$.
The log amplitudes $\gamma_j$ have independent $N(0,1)$ priors, independently of $\mu$ and the log length scales.

For $n=15$, write $\zeta_j=z_j-\log3$ and $\zeta=(\zeta_1,\ldots,\zeta_n)^{\mathsf T}$.
Integrating out $\mu$ gives covariance $\Sigma_\zeta=\sigma^2(I_n+\mathbf{1}\mathbf{1}^{\mathsf T})$ and precision
\[
\Sigma_\zeta^{-1}=\sigma^{-2}\left(I_n-\frac{\mathbf{1}\mathbf{1}^{\mathsf T}}{n+1}\right),
\] where $I_n$ is the identity matrix and $\mathbf{1}$ is the vector of ones.
Equivalently, the centred prior density is
\[
p(\zeta)\propto\exp\left[-\frac{\sum_j\zeta_j^2+\sum_{j<k}(\zeta_j-\zeta_k)^2}{2\sigma^2(n+1)}\right].
\]
This is the graph prior in \eqref{eq:graph-posterior} on the complete graph, with $a_j=w_{jk}=\{\sigma^2(n+1)\}^{-1}=1/9$.
The independent two-site analysis (I2) uses independent $N(\log3,1.125)$ priors for $z_1$ and $z_2$.

Write $\vartheta_j=(z_j,\gamma_j)$ and let $g_j(f\mid\vartheta_j)$ be the normalized Gaussian importance density obtained by expectation propagation.
For 32 independent draws $f_j^{(m)}\sim g_j(\cdot\mid\vartheta_j)$, the estimate is
\[
\widehat\ell_j(\vartheta_j,u_j)=\frac1{32}\sum_{m=1}^{32}\frac{p(y_j\mid f_j^{(m)})\,\phi_{K_j}(f_j^{(m)})}{g_j(f_j^{(m)}\mid\vartheta_j)},
\] where $p(y_j\mid f)=\prod_r\Phi(y_{jr}f_r)$, $\phi_{K_j}$ is the centred Gaussian density with covariance $K_j$, and $u_j$ contains the independent standard normal inputs used to generate the draws \citep{FilipponeGirolami2014}.
The Gaussian importance density has full support, so the estimate is nonnegative and its expectation is the marginal likelihood.
The resulting pseudo-marginal target has the covariance-parameter posterior as its marginal \citep{AndrieuRoberts2009}.

The fixed independent parameter proposal is
\[
q_j(\vartheta)=0.9\,t_5(\vartheta;\mu_j^q,1.8S_j)+0.1\,\phi_2\{\vartheta;(\log3,0)^{\mathsf T},\operatorname{diag}(1.125,1)\},
\] where $t_5$ is a normalized bivariate Student density with five degrees of freedom and the stated scale matrix, $\phi_2$ is a bivariate Gaussian density, and $\mu_j^q,S_j$ are the saved empirical mean and covariance.
These constants are fixed before sampling and supplied in \texttt{specifications/landmine/proposal.json}.
The full normalized mixture density enters the Hastings ratio.

Consider an update at site $j$ from $\vartheta_j=(\log3+x,\gamma)$ to $\vartheta'_j=(\log3+y,\gamma')$, where $x,y$ are the old and proposed centred log length scales.
For positive old and proposed estimates, define
\[
\begin{aligned} A_t={}&\log\widehat\ell'_j-\log\widehat\ell_j-\tfrac12\{(\gamma')^2-\gamma^2\} -\frac{n(y^2-x^2)}{2\sigma^2(n+1)}+\log q_j(\vartheta_j)-\log q_j(\vartheta'_j),\\ c_t={}&\frac{y-x}{\sigma^2(n+1)}, \end{aligned}
\] where $t$ is the update index and $\widehat\ell_j,\widehat\ell'_j$ are the complete old and proposed likelihood estimates.
The log acceptance ratio is $\Lambda_t=A_t+c_t\sum_{k\ne j}\zeta_k$, with neighbour values taken immediately before update $t$.

For an unresolved neighbour, take the interval hull of its latest resolved predecessor and all intervening fixed proposals for its centred log length scale.
Every accepted value is one of these proposals and every rejection retains an earlier value, so this interval contains the required historical value.
Use a singleton for a resolved historical value and write $[\underline\zeta_{kt},\overline\zeta_{kt}]$ for each neighbour interval.
Then
\[
\underline\Lambda_t=A_t+\sum_{k\ne j}\min\{c_t\underline\zeta_{kt},c_t\overline\zeta_{kt}\},\qquad \overline\Lambda_t=A_t+\sum_{k\ne j}\max\{c_t\underline\zeta_{kt},c_t\overline\zeta_{kt}\}
\] enclose the reference log acceptance ratio.
Apply the accept/reject tests in \eqref{eq:decision-certificates}, retaining the same proposal, auxiliary input and uniform threshold throughout the comparison.

Each cached record contains $(\vartheta_j,u_j,\widehat\ell_j)$, and each query resolves its own predecessor first.
Acceptance stores the complete proposed record, whereas rejection retains the complete old record.
Fix a finite update schedule, a finite ordered query list, initial records with positive likelihood estimates and common random inputs.
Work in exact arithmetic and assume that each candidate record can be generated and evaluated in finite time.
Every recursive call requests an earlier record, and each undecided comparison resolves a previously unresolved neighbour.
Once all required neighbours are resolved, the bounds coincide and the test decides the update, including equality at the threshold.
Zero proposed estimates are rejected immediately.
Given correct cached records, the historical intervals contain the reference parameters and each certificate gives the reference decision.
Induction over update times proves termination and agreement of every returned record with full reference execution.
The implementation uses outward-rounded bounds and stops if a fully resolved interval remains undecided.

\begin{figure}[tbp]
\centering
\setlength{\figwidth}{\linewidth}
\includegraphics[width=\figwidth]{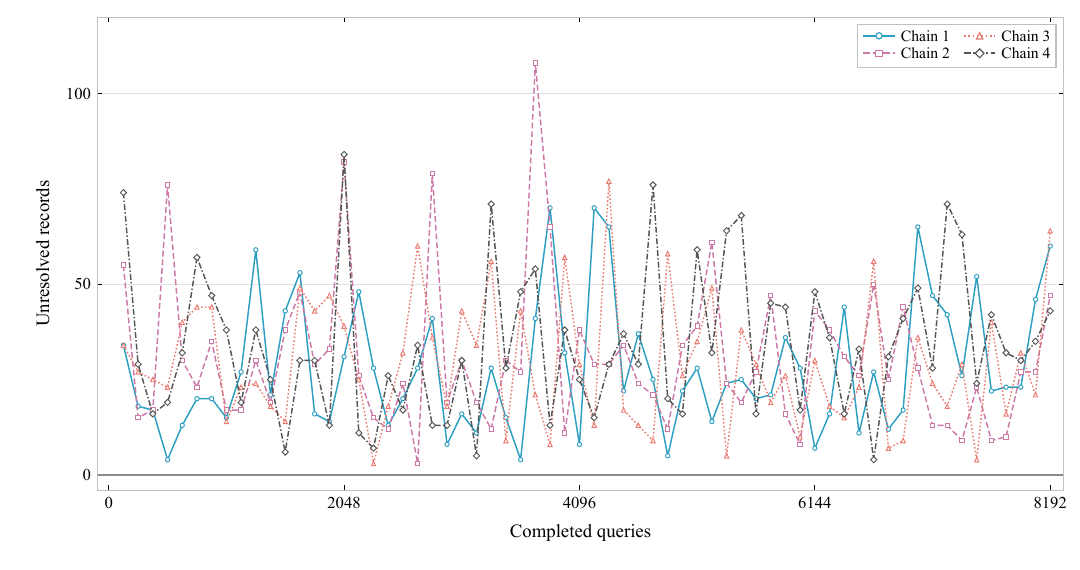}
\caption{Unresolved history during pseudo-marginal sampling for the shared 15-site Landmine model.
Sites 1 and 2 are queried jointly after every 15 random-scan updates.
The horizontal axis gives the number $q$ of completed joint queries.
The vertical axis gives $15q-C(q)$, where $C(q)$ counts the scheduled updates resolved by query $q$.
Each of the four curves follows one chain through 8192 queries, including the first 2048 used for warm-up.
Points are recorded every 128 queries and joined by line segments.}
\label{supp:landmine-history}
\figurealt{Unresolved record counts rise and fall repeatedly in all four chains. At the final query, the counts range from 43 to 64 records per chain.}
\end{figure}

For the shared 15-site model (H15), pseudo-marginal sampling uses four chains, 8192 nominal sweeps per chain and 15 random-scan updates per sweep.
The first 2048 sweeps are discarded, leaving 6144 joint frames per chain.
Figure~\ref{supp:landmine-history} shows the number of unresolved records after successive queries in the four pseudo-marginal chains.
In chain~4, this number falls from 84 to 11 between queries 2048 and 2176.
The executor resolves all 84 earlier records during this interval.
The 11 unresolved records at query~2176 belong to updates after query~2048.

The ancillary augmentation sampler writes the latent vector as
$f_j=L_j\nu_j+b_j\mathbf{1}$, where $\nu_j$ has a standard normal
prior, $b_j\sim N(0,4)$ and $L_jL_j^{\mathsf T}$ is the covariance
without the constant intercept term.
Ten pairs of elliptical slice updates of $\nu_j$ and $b_j$ precede a joint symmetric Metropolis--Hastings update of $(z_j,\gamma_j)$ \citep{MurrayAdams2010,MurrayAdamsMacKay2010}.
Sites are visited systematically.
We tune the parameter proposal scales in a separate pilot, with one scale for each site in each chain.
The pilot uses 256 tuning sweeps with a target acceptance rate of 0.234, followed by 2048 sweeps with the scales fixed.
The four reported chains start from the final pilot states and use these fixed scales.
Pilot draws are excluded from the reported samples.
Each chain is run for 49152 sweeps and extended by 32768 sweeps with the same transition kernel.
Discarding the first 8192 of the combined 81920 sweeps leaves 73728 frames per chain.
I2 uses four chains of 8192 systematic two-site sweeps, with 2048
discarded. It uses the same fixed parameter proposals with the I2
prior in the acceptance ratio.

\subsection{Diagnostics and Monte Carlo error}
\label{supp:diagnostics}
We use the rank-normalized and folded split-chain diagnostics of
\citet{VehtariEtAl2021}. Each chain is split into two equal halves.
Tied values receive their average rank.
We map each pooled rank $r$ to the normal score $\Phi^{-1}\{(r-3/8)/(S+1/4)\}$, where $S$ is the number of retained draws across the split chains and $\Phi$ is the standard normal distribution function. The reported $\widehat R$
is the larger value from the ranked draws and their ranked absolute
deviations from the pooled median.
Bulk effective sample size uses the rank-normalized draws. Tail
effective sample size is the smaller estimate for indicators at the
5\% and 95\% empirical quantiles. The autocorrelation estimate uses
initial positive and monotone paired sums. The implementation caps
effective sample size at the number of saved draws.

For a mean, the tomography analysis reports the larger of the raw-scale
effective-sample-size estimate of standard error and the standard error
of the four chain means. Landmine additionally includes non-overlapping
batch-mean estimates at batch lengths 128 and 256, taking the maximum
of all four estimates. This rule also applies to the direction indicator.
For a quantile probability $p$, we estimate the effective sample size $n_e$ from the indicators of retained draws at or below the empirical $p$-quantile.
Following \citet{VehtariEtAl2021}, we compute the 16\% and 84\% quantiles of $\mathrm{Beta}\{n_ep+1,n_e(1-p)+1\}$.
We map these two probabilities to the scale of the retained draws.
Tomography uses the bounding order statistics, and Landmine uses linear empirical-quantile interpolation.
Half the difference between the upper and lower mapped values estimates the Monte Carlo standard error.
These are estimated simulation errors, conditional on the chosen statistical and numerical targets.

Tables~\ref{supp:pet-diag} and~\ref{supp:landmine-diag} report chain diagnostics for tomography and Landmine, respectively.
Table~\ref{supp:landmine-errors} reports estimated Monte Carlo standard errors for the Landmine contrast $d$.

\begin{table}[tbp]
\tbl{Tomography chain diagnostics}{%
\begin{tabular}{lrrr}
Quantity & $\widehat R$ & Bulk ESS & Tail ESS\\[4pt]
Conditional regional mean & 1.007 & 535 & 961\\
Regional draw & 1.000 & 7113 & 7014\\
Two-tissue fraction & 1.009 & 497 & 991\\
\end{tabular}}
\label{supp:pet-diag}
\begin{tabnote}
ESS denotes effective sample size.
The two-tissue fraction is the proportion of regional voxels assigned to the two-tissue compartment model.
Each of the four chains retains 1792 queries.
For the regional average of $V_T$, the 95\% equal-tailed posterior interval has estimated Monte Carlo standard errors of $1.93\times10^{-4}$ and $1.57\times10^{-4}\,\mathrm{mL}\,\mathrm{cm}^{-3}$ at its lower and upper endpoints, respectively.
These standard errors are $0.87\%$ and $0.71\%$ of the interval width, respectively.
\tablealt{Bulk effective sample sizes are 535 for the conditional regional mean, 7113 for the regional draw and 497 for the two-tissue fraction. All three convergence statistics are close to one.}
\end{tabnote}
\end{table}
\begin{table}[tbp]
\tbl{Landmine chain diagnostics}{%
\begin{tabular}{lllrrr}
Model & Sampler & Quantity & $\widehat R$ & Bulk ESS & Tail ESS\\[4pt]
H15 & PM & $z_1$ & 1.000 & 7603 & 7534\\
H15 & PM & $z_2$ & 1.000 & 7505 & 8072\\
H15 & PM & $d$ & 1.000 & 8273 & 11546\\
H15 & AA & $z_1$ & 1.001 & 2284 & 4064\\
H15 & AA & $z_2$ & 1.002 & 1996 & 4534\\
H15 & AA & $d$ & 1.001 & 2384 & 4745\\
I2 & PM & $z_1$ & 1.000 & 12451 & 14376\\
I2 & PM & $z_2$ & 1.000 & 14028 & 14718\\
I2 & PM & $d$ & 1.000 & 13177 & 16449\\
\end{tabular}}
\label{supp:landmine-diag}
\begin{tabnote}
H15 denotes the shared 15-site model and I2 the independent two-site model.
PM and AA denote pseudo-marginal and ancillary augmentation sampling, respectively.
ESS denotes effective sample size.
The contrast $d=z_1-z_2$ compares the log length scales of sites 1 and 2.
All analyses use four chains.
Each H15 PM or I2 PM chain retains 6144 frames; each H15 AA chain retains 73728.
\tablealt{Across both models, all reported convergence statistics are between 1.000 and 1.002. In the shared model, pseudo-marginal sampling gives larger bulk and tail effective sample sizes for all three quantities.}
\end{tabnote}
\end{table}
\begin{table}[tbp]
\tbl{Landmine Monte Carlo standard errors}{%
\begin{tabular}{llrrrr}
Model & Sampler & Mean & Direction probability & Lower endpoint & Upper endpoint\\[4pt]
H15 & PM & 0.00868 & 0.00513 & 0.0188 & 0.0212\\
H15 & AA & 0.0147 & 0.00836 & 0.0180 & 0.0254\\
I2 & PM & 0.00971 & 0.00648 & 0.0139 & 0.0241\\
\end{tabular}}
\label{supp:landmine-errors}
\begin{tabnote}
H15 denotes the shared 15-site model and I2 the independent two-site model.
PM and AA denote pseudo-marginal and ancillary augmentation sampling, respectively.
The contrast $d=z_1-z_2$ compares the log length scales of sites 1 and 2.
The columns give estimated Monte Carlo standard errors for its posterior mean, probability of being positive and two endpoints of its 95\% equal-tailed posterior interval.
For each H15 summary, we estimate the Monte Carlo standard error of the difference between the PM and AA estimates by taking the square root of the sum of their squared standard errors.
\tablealt{Across the three analyses, Monte Carlo standard errors range from 0.00513 to 0.00836 for the probability of a positive site contrast and from 0.0139 to 0.0254 for the interval endpoints.}
\end{tabnote}
\end{table}

For Landmine, we use thresholds of $\widehat R\le1.01$ and bulk and tail effective sample sizes of at least 400 for each of $z_1,z_2,d$.
We require estimated Monte Carlo standard errors of at most $0.02$ for the posterior mean of $d$ and the posterior probability that $d>0$.
The estimated Monte Carlo standard error for each interval endpoint of $d$ must be at most $0.03$. A constant direction indicator in any chain is also
flagged. In tomography, we check $\widehat R<1.01$ and a bulk effective sample size of at least 400 for the conditional regional mean, regional draw and two-tissue fraction.
The estimated Monte Carlo standard error of the posterior mean of the regional average must be at most $0.5\%$ of the estimated mean.
For each endpoint of the 95\% equal-tailed interval, we require an indicator effective sample size of at least 400 and an estimated Monte Carlo standard error of at most $5\%$ of the interval width. These checks do not constitute
a bound on distance from stationarity.

\section{Accompanying records}
\label{supp:records}
The companion source package contains the fixed model configurations,
the code excerpts defining numerical rules and diagnostics, the recorded
timings, and a file-identity manifest. These files identify the quantities
used in this supplement. They are a record of the reported analyses;
the package does not launch new posterior simulation.
The repository paths and source hashes in the manifest permit matching
them to the complete project before a public reproducibility release.
The public dataset identifiers are given in the references.

\end{document}